\documentclass[aps,prb,preprint,floatfix,showpacs]{revtex4-1}
\usepackage{graphicx}
\usepackage{amsmath}
\usepackage{amssymb}
\usepackage{dcolumn}
\graphicspath{{./}}
\providecommand{\e}[1]{\ensuremath{\times 10^{#1}}}
\providecommand{\rate}[2]{\lambda_{\text{#1-#2}}}

\providecommand{\pa}[1]{\ensuremath{\rho_{\rm{#1}}}}
\newcommand*{\kB}[0]{k_{\text{B}}}
\newcommand*{\kT}[0]{\kB T}
\begin{document}
\title{Light element diffusion in Mg using first principles calculations: Anisotropy and elastodiffusion}
\author{Ravi Agarwal}
\author{Dallas R. Trinkle}
\email{dtrinkle@illinois.edu}
\affiliation{Department of Materials Science and Engineering, University of Illinois at Urbana-Champaign, Urbana, IL 61801}
\date{\today}

\begin{abstract}
The light elemental solutes B, C, N, and O can penetrate the surface of Mg alloys and diffuse during heat treatment or high temperature application, forming undesirable compounds. We investigate the diffusion of these solutes by determining their stable interstitial sites and the inter-penetrating network formed by these sites. We use density functional theory (DFT) to calculate the site energies, migration barriers, and attempt frequencies for these networks to inform our analytical model for bulk diffusion. Due to the nature of the networks, O diffuses isotropically, while B, C, and N diffuse anisotropically. We compute the elastodiffusion tensor which quantifies changes in diffusivity due to small strains that perturb the diffusion network geometry and the migration barriers. The DFT-computed elastic dipole tensor which quantifies the change in site energies and migration barriers due to small strains is used as an input to determine the elastodiffusion tensor. We employ the elastodiffusion tensor to determine the effect of thermal strains on interstitial diffusion and find that B, C, and N diffusivity increases on crystal expansion, while O diffusivity decreases. From the elastodiffusion and compliance tensors we calculate the activation volume of diffusion and find that it is positive and anisotropic for B, C and N diffusion, whereas it is negative and isotropic for O diffusion.
\end{abstract}
\pacs{66.10.C-, 66.10.cg, 66.30.J-, 66.30.Ny}

\maketitle

\section{Introduction}
Magnesium and its alloys have found increased application in the automotive industry due to their higher strength-to-weight ratio than steel and aluminum alloys, which reduces vehicle weight leading to increase in fuel efficiency\cite{Pollock2010,MgSpringer2006,Joost2012}. Mg alloys interact with the surrounding gaseous atmosphere during their application which can lead to the penetration of light impurity elements. These impurities can also get introduced due to interaction with reactive gases during heat treatment, leading to the formation of oxide layers on the surface or precipitates at grain boundaries which can be detrimental to strength\cite{Fromm1980,MgSpringer2006}. Experiments have shown that O, C and N can react with Mg to form oxides, carbides and nitrides\cite{MgSpringer2006}. Boron is used for Fe removal during Mg processing\cite{MgSpringer2006}, but a small amount of B may be retained as an impurity. The penetration of these impurities into bulk is governed by thermally activated processes and a detailed study of their diffusion mechanisms can provide insights that may help to mitigate them.

There have been few theoretical studies on the behavior of light elements in hcp metals. Wu~\textit{et al.} studied the influence of substitutional B, C, N and O on the stacking faults and surfaces of Mg\cite{Wu2013} using density functional theory (DFT). All four elements reduce the unstable stacking fault energy and surface energy of Mg and enhance the ductility according to the Rice criterion, with O having the largest impact\cite{Wu2013}. Atomisitic studies of light elements in hcp metals---O in $\alpha$-Ti\cite{Bertin1980,Wu2011}, O and N diffusion in $\alpha$-Hf\cite{Hara2014}, and O in multiple hcp metals\cite{Wu2016}---modeled the diffusion of solutes through the networks formed by interstitial sites. However, a theoretical or experimental study of interstitial diffusion in Mg is absent except for the limited experimental data for C diffusion\cite{Zotov1976}. 

We analyze the diffusion of B, C, N and O in the dilute limit in hcp Mg using DFT calculations to inform an analytical diffusion model\cite{TrinkleElastodiffusivity2016,OnsagerCalc}. We also study the changes in diffusivities due to strain from thermal expansion. Section II details the DFT parameters used to determine the energetics of interstitial sites and the migration barriers between them. Section III lays out the inputs for the  diffusion model: probabilities of occupying sites, connectivity networks between these sites and the transition rates for these networks. We derive analytical expressions for interstitial diffusivity in hcp crystals and apply them to diffusion of B, C, N and O in Mg. We find that the O diffusion is isotropic while B, C, and N diffusion is anisotropic. Section IV discusses the elastic dipole tensors of solutes at interstitial sites and transition states, which determine the changes in the transition energetics of solutes due to small strains. Section V defines the elastodiffusion tensor\cite{Dederichs1978,Savino1987,Woo2000,TrinkleElastodiffusivity2016}, which quantifies the effect of small strains on diffusivity and discusses the sign inversion behavior of elastodiffusion components with temperature. We find that the activation volume of O diffusion is negative which leads to an increase in O diffusion under hydrostatic pressure. We also find that the diffusivity of O decreases with thermal expansion while the diffusivity of B, C and N increases. 

\section{Computational details}

We perform the DFT calculations using the Vienna ab-initio simulation package \textsc{vasp}\cite{Kresse1996} which is based on plane wave basis sets. The projector-augmented wave psuedopotentials\cite{Blochl1994a} generated by Kresse\cite{Kresse1999} describe the nuclei and the valence electrons of solutes and Mg atoms. The solute atoms B, C, N, and O are described by [He] core with 3, 4, 5 and 6 valence electrons respectively. We use the [Ne] core with 2 valence electrons for Mg instead of the [Be] core with 8 valence electrons because the energies computed using either choice of psuedopotential differ by less than 20 meV. Electron exchange and correlation is treated using the PBE\cite{Perdew1996} generalized gradient approximation. We use a $4\times4\times3$ (96 atoms) supercell of Mg atoms with a $6\times6\times6$ Monkhorst-Pack $k$-point mesh to sample the Brillouin zone. Methfessel-Paxton smearing\cite{Methfessel1989} is used with energy width of 0.25 eV to integrate the density of states; the k-point density and smearing width are based on convergence of the DOS compared with tetrahedron integration. A plane wave energy cutoff of 500 eV is required to give an energy convergence of less than 1 meV/atom. All the atoms are relaxed using a conjugate gradient method until each force is less than 5 meV/\AA. The Mg unit cell has a hexagonal close packed (HCP) crystal structure with DFT calculated lattice parameters of $a=3.189\text{ \AA}$ and $c/a$ ratio of 1.627 which agree well with values reported from experiments, $a= 3.19\text{ \AA}$ and $c/a=1.62$\cite{Friis2003}.

We use DFT to calculate the energy of solutes at various sites and use the climbing-image nudged elastic band (CNEB)\cite{Henkelman2000} method to locate the transition states between the sites. The site (or solution) energy $E_{\alpha}$ of a solute X at an interstitial site $\alpha$  is the difference between the energy of a Mg supercell containing solute X at site $\alpha$, $E(\text{Mg}_{96}+\text{X}_{1}^{\alpha})$, and the energy of a pure Mg supercell, $E(\text{Mg}_{96})$, 
\begin{equation}
\label{eq:formation}
E_{\alpha}=E(\text{Mg}_{96}+\text{X}_{1}^{\alpha})-E(\text{Mg}_{96}).
\end{equation}
We also determine the site energy for a solute X as a substitutional defect, $E_{\text{sub}}$,
\begin{equation}
\label{eq:subs_formation}
E_{\text{sub}}=E(\text{Mg}_{95}+\text{X}_{1}^{\text{sub}})-\frac{95}{96} E(\text{Mg}_{96})
\end{equation}
where $E(\text{Mg}_{95}+\text{X}_{1}^{\text{sub}})$ is the energy of supercell where one of the Mg atoms is substituted by a solute atom X. Both the interstitial site energy $E_{\alpha}$ and the substitutional site energy $E_{\text{sub}}$ for solute X are referenced to its elemental state. The energy differences $\Delta E=E_{\alpha}-E_{\text{sub}}$ for the solutes B, C, N and O are --1.48, --3.23, --4.34 and --4.19 eV, where $\alpha$ is the interstitial site with the lowest energy, and is independent of the reference state for the solutes. Since, the energies of interstitial sites are lower than the substitutional site, these solutes are likely to diffuse through networks of interstitial sites. We use CNEB with one image\cite{Henkelman2000} to locate the transition state between two interstitial sites. Similar to Eq.~\ref{eq:formation}, the energy $E_{\alpha \textnormal{-}\beta}$ of the transition state between site $\alpha$ to site $\beta$ is referenced to the elemental state of X
\begin{equation}
\label{eq:trans_formation}
E_{\alpha\textnormal{-}\beta}=E_{\alpha\textnormal{-}\beta}(\text{Mg}_{96}+\text{X}_{1})-E(\text{Mg}_{96})
\end{equation}
where $E_{\alpha\textnormal{-}\beta}(\text{Mg}_{96}+\text{X}_{1})$ is the energy at the transition state obtained from a CNEB calculation. We report the interstitial site energies and the transition state energies relative to the interstitial site with the lowest energy, which is independent of the reference state for the solutes. 

\section{Diffusion model}
We calculate the occupation probabilities at interstitial sites and transition rates for diffusion pathways from DFT-computed site energies, transition state energies, and vibrational frequencies. The probability $\pa{\alpha}$ of a solute occupying a particular site $\alpha$ at temperature $T$ is
\begin{equation}
\pa{\alpha}=\frac{\nu_{\alpha}^{*}\cdot\textnormal{exp}(-E_{\alpha}/\kT)}{\sum_{\beta}\nu_{\beta}^{*}\cdot\textnormal{exp}(-E_{\beta}/\kT)},
\label{eq:probability}
\end{equation}
where $\kB$ is the Boltzmann constant, in the denominator is the normalization constant summed over all the interstitial sites in the unit cell and $\nu_{\alpha}^{*}$ is the site prefactor proportional to the Arrhenius factor for formation entropy of site $\alpha$, $\exp\left(S_\alpha/\kB\right)$, calculated from the vibrational frequencies 
\begin{equation}
\nu_{\alpha}^{*}=\frac{1}{\prod_{p=1}^{3}\nu_{\alpha,p}}.
\end{equation}
This expression ignores interstitial-interstitial interaction, and is exact in the dilute concentration limit. We compute the vibrational frequencies of a state using the one atom approximation by diagonalizing the dynamical matrices corresponding to the interstitial atom.%
\footnote{This approximation introduces at most a 40\%\ error in the attempt frequencies; the error is estimated by comparing with a large Mg supercell using bulk force constants, and introducing the interstitial-Mg force constants from the finite displacement calculations.}
The dynamical matrices are obtained from the forces induced on interstitial atoms by small displacements ($\pm 0.01$ \AA) from their equilibrium positions, while keeping the other atoms fixed. From transition state theory, the rate $\rate{$\alpha$}{$\beta$}$ for a solute to transition from site $\alpha$ to site $\beta$ at temperature $T$ is
\begin{equation}
\rate{$\alpha$}{$\beta$}=\nu_{\alpha\textnormal{-}\beta}^{*}\cdot\textnormal{exp}(-(E_{\alpha\textnormal{-}\beta}-E_{\alpha})/\kT).
\label{eq:rates}
\end{equation}
The attempt frequency $\nu_{\alpha\textnormal{-}\beta}^{*}$ for the $\alpha$ to $\beta$ transition is calculated using the Vineyard expression\cite{Vineyard1957}, which is the product of vibrational frequencies $\nu_{\alpha,p}$ at the initial site $\alpha$ divided by the product of real vibrational frequencies $\nu_{\alpha\textnormal{-}\beta,q}$ at the transition state
\begin{equation}
\nu_{\alpha\textnormal{-}\beta}^{*}=\frac{\prod_{p=1}^{3}\nu_{\alpha,p}}{\prod_{q=1}^{2}\nu_{\alpha\textnormal{-}\beta,q}}.
\end{equation} 
At equilibrium, the transition between site $\alpha$ and site $\beta$ obeys detailed balance 
\begin{equation}
\pa{\alpha}\cdot\rate{$\alpha$}{$\beta$}=\pa{\beta}\cdot\rate{$\beta$}{$\alpha$}.
\end{equation}

\begin{figure}
\includegraphics[width=0.5\textwidth]{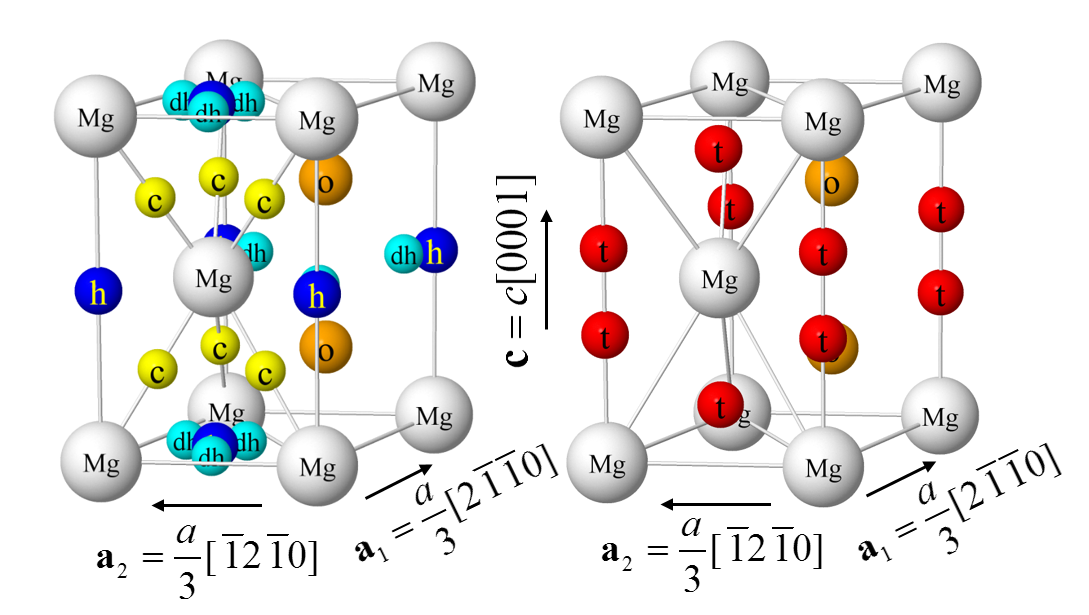}

\caption{(color online) Positions of interstitial sites in the unit cell of hcp Mg. The octahedral (o, orange), tetrahedral (t, red), hexahedral (h, blue), distorted hexahedral (dh, cyan), and crowdion (c, yellow) interstitial sites are shown relative to host Mg atoms (Mg, white). In an hcp unit cell, there are two o, two h, four t, six c and six dh sites. The transitions between stable interstitial sites determine the possible diffusion pathways. The unit cell vectors $\mathbf{a}_1$ and $\mathbf{a}_2$ form the basal plane (0001) and the vector $\mathbf{c}$ (also referred as the $c$-axis ) is perpendicular to it.}
\label{fig:Unitcell}
\end{figure}

Figure~\ref{fig:Unitcell} shows the newly found distorted hexahedral dh site in Mg along with the other interstitial sites (h, t, c, o) which have been discussed previously for O in $\alpha$-Ti\cite{Wu2011}. The dh site is stable for B and C, and is located between two nearest Mg atoms in the basal plane with a displacement of 0.17 \AA\ for B and 0.40 \AA\ for C towards the nearest hexahedral h site. The h site has three basal Mg neighbors and two other Mg neighbors located directly above and below it, which are further away. The four-atom coordinated tetrahedral t site is stable for O and lies 0.65 \AA\ along the $c$ direction from an basal plane containing three of its Mg neighbors. The six-atom coordinated octahedral o site is stable for all four solutes. The six-atom coordinated non-basal crowdion c with lower symmetry than o site has two nearest neighboring Mg atoms lying in adjacent basal planes which get displaced away from the c site while the other four neighbors lying further apart get displaced towards the c site on relaxation. The c site is stable for C and N but unstable for B and O.   

\begin{figure}
  \begin{minipage}[t]{0.24\textwidth}
    \includegraphics[width=\textwidth]{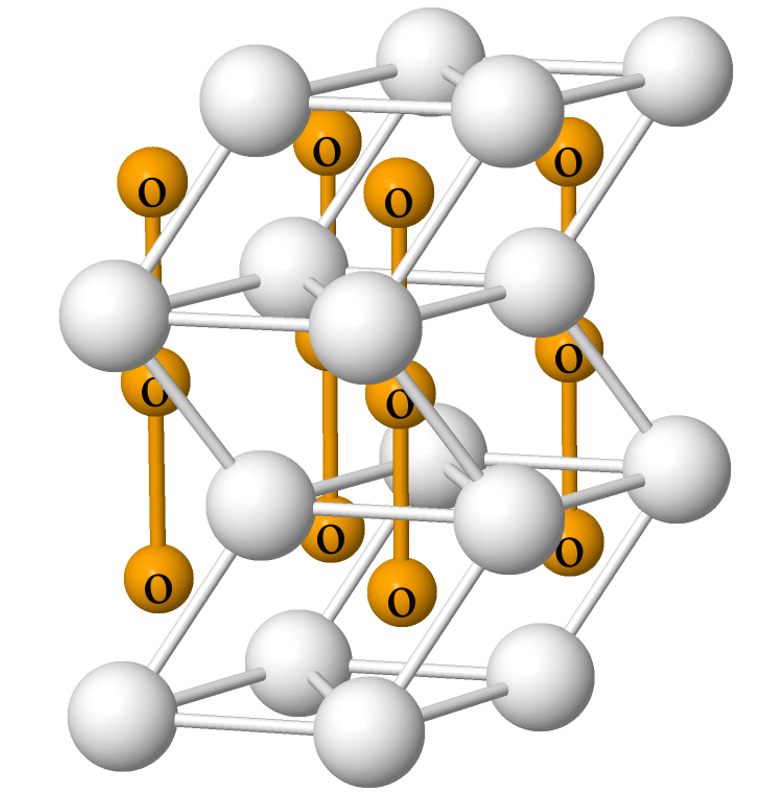}
    {o-o($c$)}
  \end{minipage}
  \begin{minipage}[t]{0.24\textwidth}
    \includegraphics[width=\textwidth]{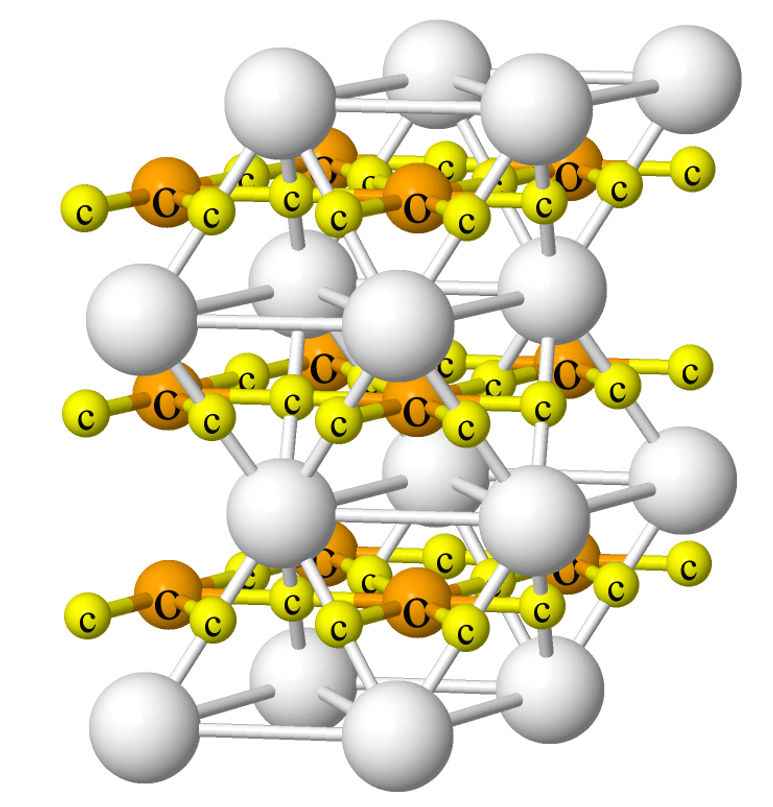}
    {o-o(b), o-c}
  \end{minipage}
  \begin{minipage}[t]{0.24\textwidth}
    \includegraphics[width=\textwidth]{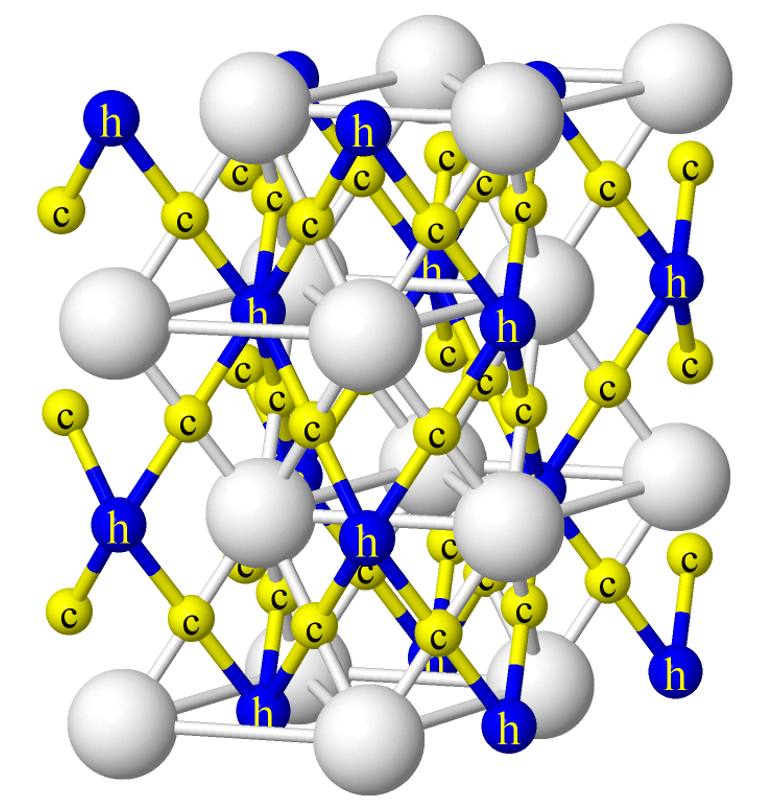}
    {h-c}
  \end{minipage}
  \begin{minipage}[t]{0.24\textwidth}
    \includegraphics[width=\textwidth]{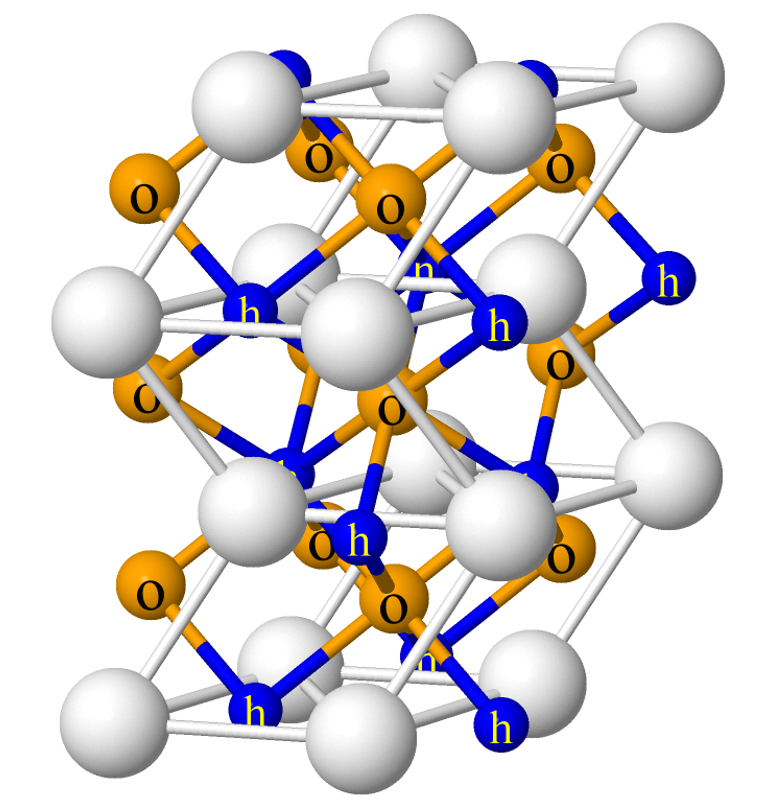}
    {o-h}
  \end{minipage}

  \begin{minipage}[t]{0.24\textwidth}
    \includegraphics[width=\textwidth]{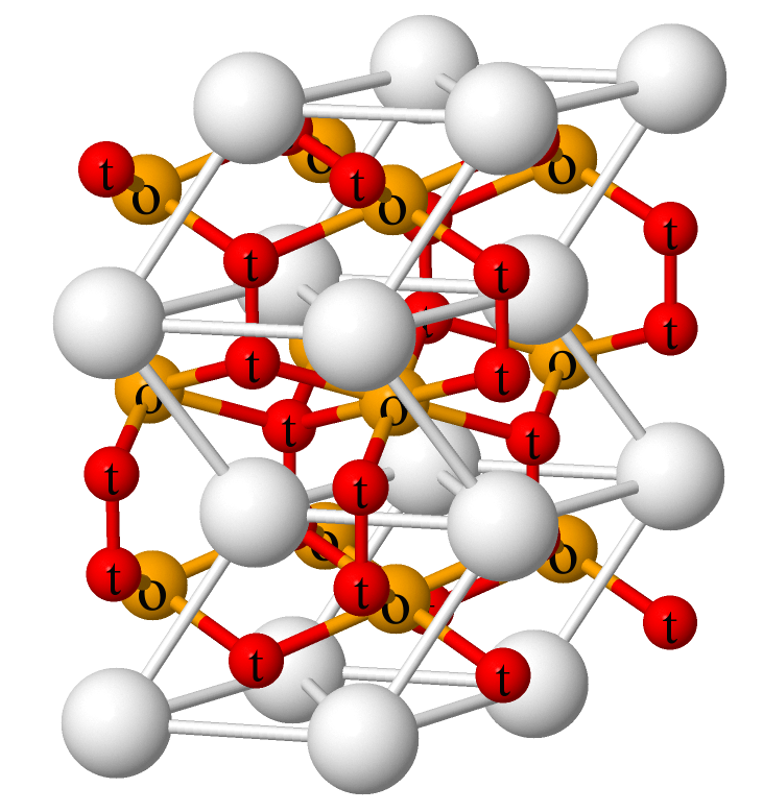}
    {t-t, t-o}
  \end{minipage}
  \begin{minipage}[t]{0.24\textwidth}
    \includegraphics[width=\textwidth]{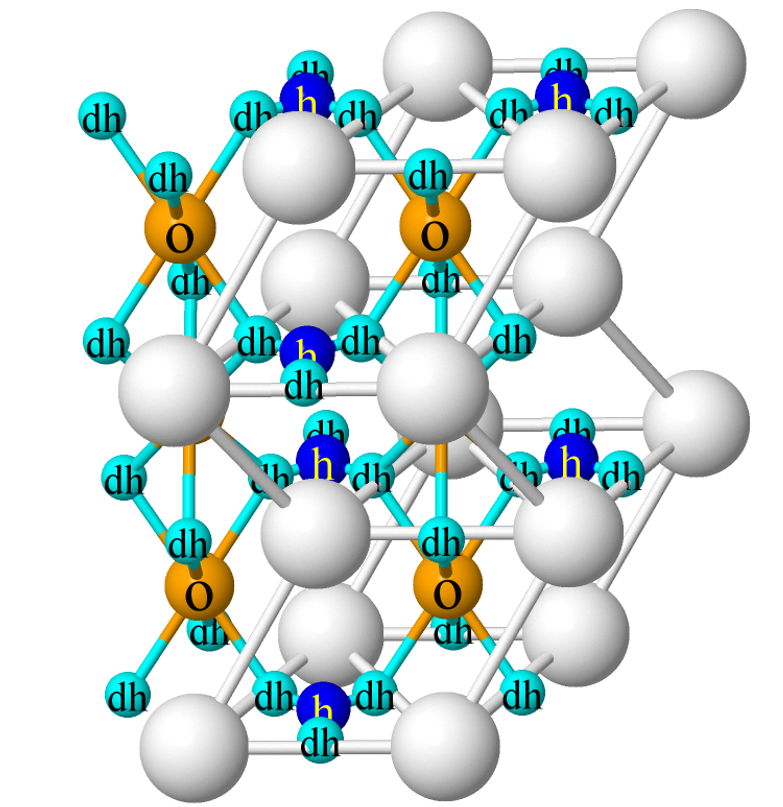}
    {o-dh, dh-h}
  \end{minipage}
  \begin{minipage}[t]{0.24\textwidth}
    \includegraphics[width=\textwidth]{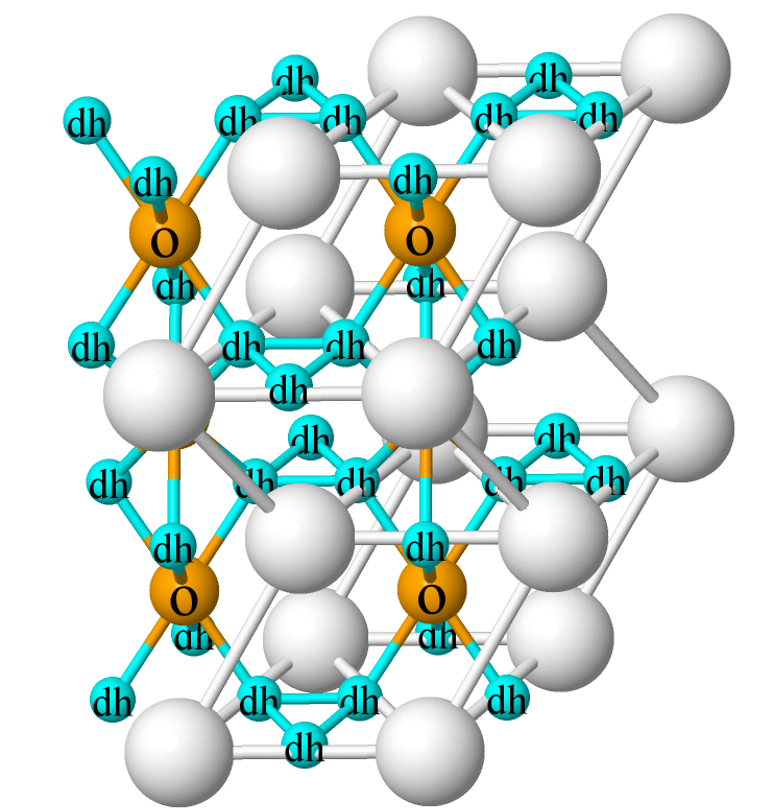}
    {o-dh, dh-dh}
  \end{minipage}
  \caption{(color online) 
	Interstitial sites and site-to-site connectivity in hcp crystals. Connections between two neighboring sites form diffusion pathways which are shown as lines colored corresponding to the colors of the interstitial sites. For example, o-o($c$) and o-o(b) are octahedral site-to-octahedral site diffusion pathways along the $c$-axis and in the basal plane of hcp Mg, respectively. The diffusion pathways shown in the top row are un-correlated, while  correlated diffusion pathways are shown in the bottom row. These correlated pathways are the combined connections formed among o and t sites, o, dh and h sites, and o and dh sites. In the last two figures of the bottom row, the $c$-axis is tilted and the cell is rotated counter-clockwise around the $c$-axis for better visibility of connections and sites.
}
\label{fig:Network}
\end{figure}

Figure~\ref{fig:Network} shows the possible diffusion networks between interstitial sites for hcp systems, which are inputs to our diffusion model\cite{TrinkleElastodiffusivity2016,OnsagerCalc}. A solute at a o site can jump to the following neighboring sites: two o sites lying above and below along the $c$-axis with transition rate $\rate{o}{o($c$)}$; six o sites lying in the same basal plane with $\rate{o}{o(b)}$ in cases where the c site is unstable ; six neighboring c sites with $\rate{o}{c}$; six h sites with $\rate{o}{h}$; six t sites with $\rate{o}{t}$ and six dh sites with $\rate{o}{dh}$. A solute at a h site can jump to: six o sites with $\rate{h}{o}$; six c sites with $\rate{h}{c}$ and three dh sites lying in the same basal plane with $\rate{h}{dh}$. The c site is between two h sites which lie in adjacent basal planes and also between two o sites in the same basal plane. A solute from a c site can jump to those neighboring o and h sites with $\rate{c}{o}$ and $\rate{c}{h}$. A solute at a t site can jump to three neighboring o sites which are all lying either above or below the t site with $\rate{t}{o}$, and to one neighboring t site lying either above or below with $\rate{t}{t}$. A solute at a dh site can jump to one neighboring h site with $\rate{dh}{h}$ and to two nearest dh sites in the same basal plane with $\rate{dh}{dh}$. 

\begin{figure}
  \begin{minipage}[t]{0.24\textwidth}
    \includegraphics[width=\textwidth]{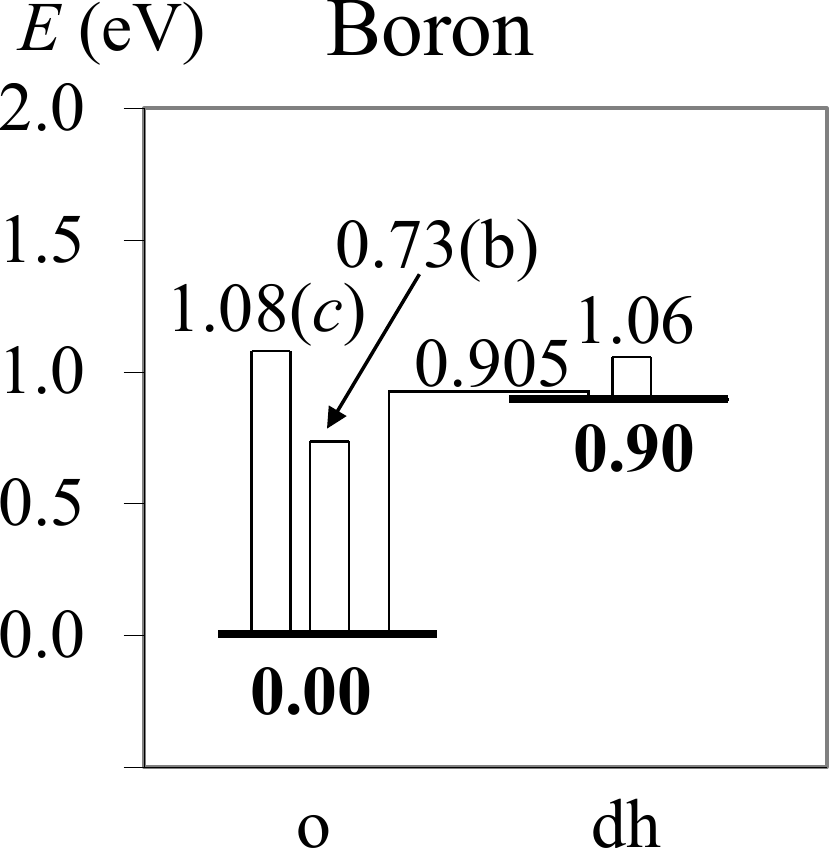}
  \end{minipage}
  \begin{minipage}[t]{0.24\textwidth}
    \includegraphics[width=\textwidth]{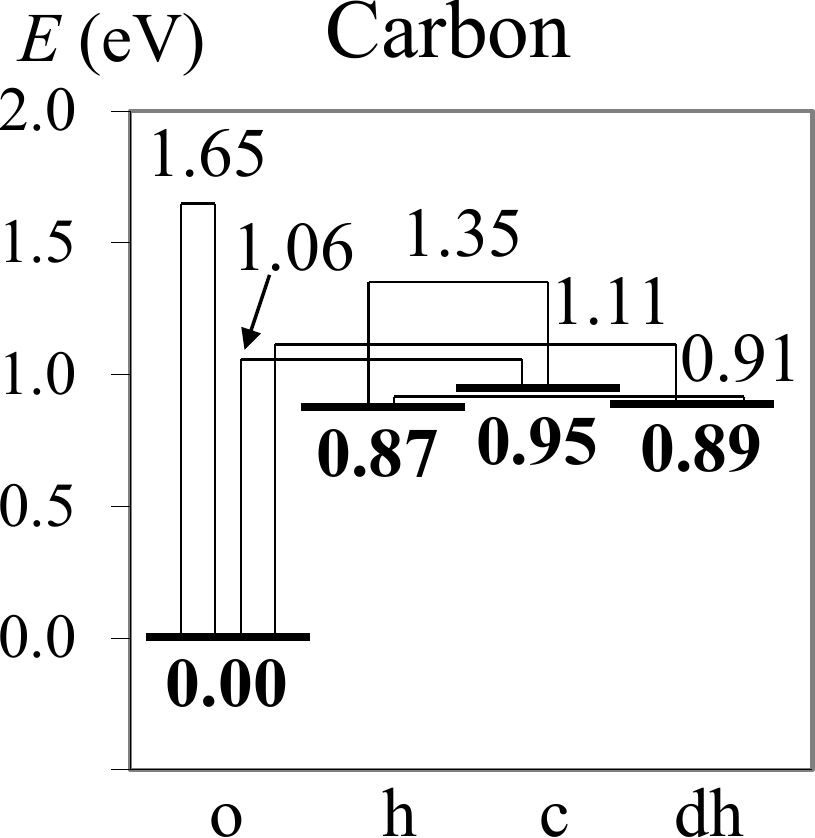}
  \end{minipage}

  \begin{minipage}[t]{0.24\textwidth}
    \includegraphics[width=\textwidth]{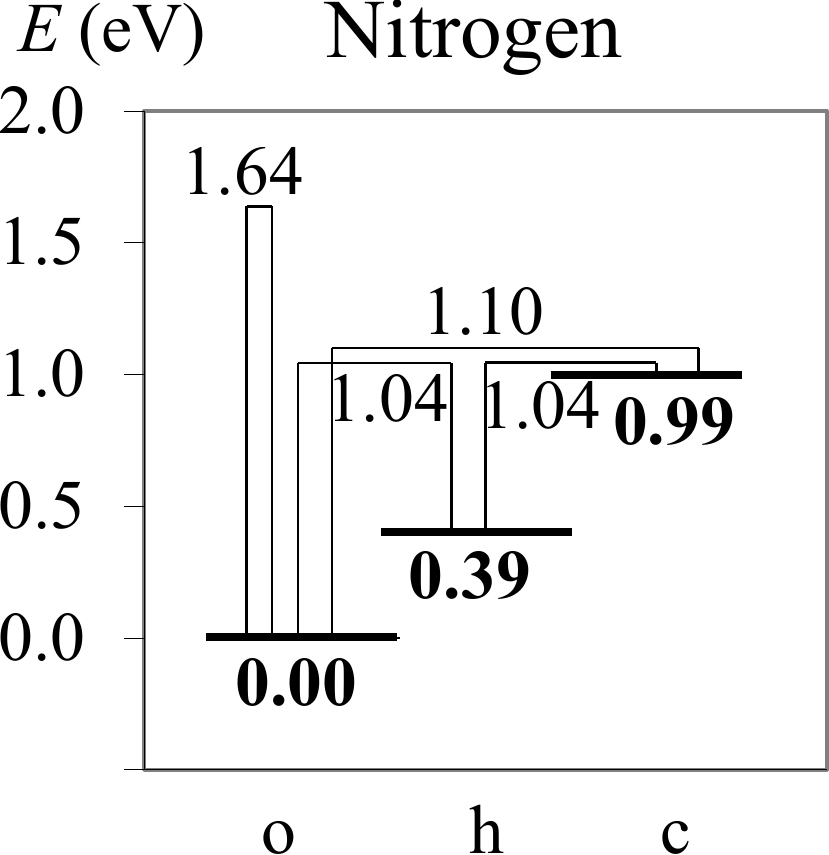}
  \end{minipage}
  \begin{minipage}[t]{0.24\textwidth}
    \includegraphics[width=\textwidth]{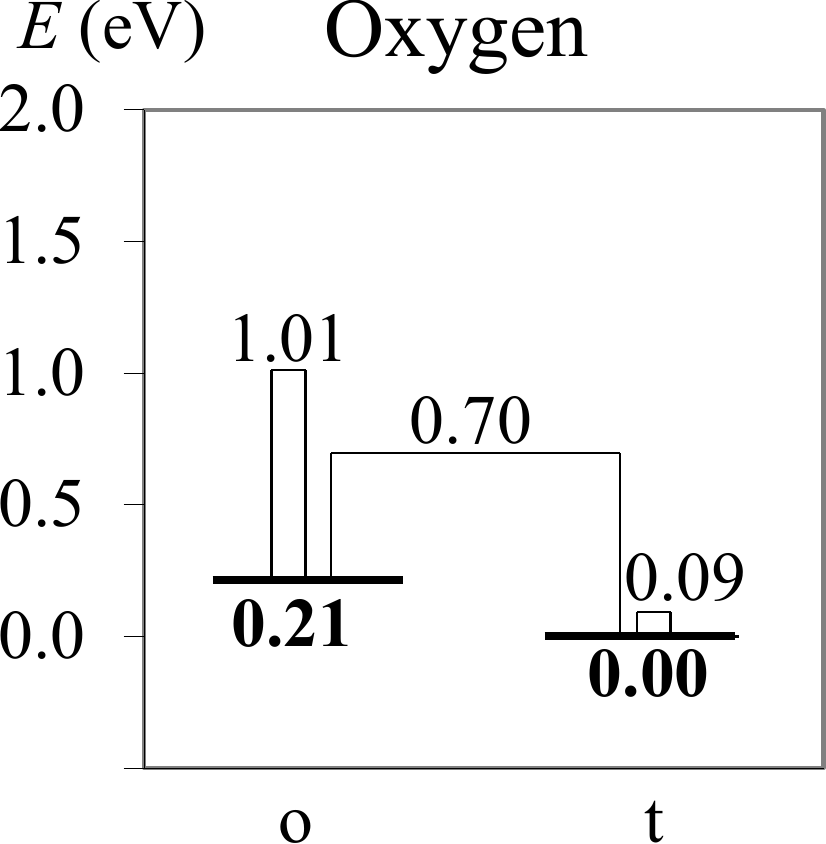}
  \end{minipage}

	\caption{Energetics of stable sites and the transition states between them, relative to the lowest-energy interstitial site for B, C, N, and O solutes in Mg. Interstitial sites are marked on the horizontal axis, and their relative site energies are shown in bold below the thick horizontal base lines. Thin lines from one site to another (or the same) site denote transitions, and the associated number is the energy at the transition state between those two sites. For example, in the case of B, the o site is the lowest energy site and the energy of the metastable dh site relative to it is 0.90 eV. Thin lines starting and ending from the thick base line of o denotes the o-o transition. The associated transition state energies in eV are 1.08($c$) for the transition along the $c$-axis and 0.73(b) for the transition in the basal plane.
}
\label{fig:Barrier}
\end{figure}

Figure~\ref{fig:Barrier} shows the energies for the interstitial sites and the transition states of active diffusion pathways for all four solutes. Active diffusion pathways for a solute are determined by its set of stable sites. The set of stable sites for B is \{o, dh\}, for C it is \{o, h, c, dh\}, for N it is \{o, h, c\} and for O it is \{o, t\}. All DFT energies are relative to the lowest-energy site which is the ground state.%
\footnote{Following Varvenne~\textit{et al.}\cite{Varvenne2013}, we can estimate the finite-size error from using a $4\times4\times3$ cell from the elastic dipoles (c.f., Table~\ref{tab:Sitedipole}) and elastic constants. The largest (estimated) error in site energies---relative to the ground state---are 80 meV for B (dh), 60 meV for C (dh), 20 meV for N (h), and 3 meV for O (t).}
The o site is the ground state for B, C and N, while the t site is the ground state for O. The transition between two sites is shown as a line connection and the associated value is the transition state energy. For example, in the case of O, t is the ground state and o is metastable with energy 0.21 eV. The active diffusion pathways for O (refer to Fig.~\ref{fig:Network}) are o-o, t-t (both along the $c$-axis), and t-o with transition state energies of 1.01, 0.09 and 0.7 eV respectively. Since there is no direct o-o (b) jump in the basal plane---which would pass through the unstable c site---basal diffusion occurs by combining o-t and t-o jumps. 

\begin{table}
\caption{Analytical expressions for interstitial solute diffusivity in the basal plane ($D_{\text{b}}$) and along the $c$-axis ($D_{c}$) through the network formed by interstitial sites in the hcp crystal. These expressions are functions of transition rates ($\lambda$) between interstitial sites and the occupation probability of each type of interstitial site. The occupation probability of each type of site is the product of $\rho$ (from Eq. \ref{eq:probability}) and its multiplicity in the unit cell. The occupation probability for any o, h, t, dh and c site is $2\pa{o}$, $2\pa{h}$, $4\pa{t}$, $6\pa{dh}$ and $6\pa{c}$, respectively. These analytical expressions for diffusivity are valid for any interstitial solute diffusing in an hcp crystal with lattice parameters $a$ and $c$ and having a set of stable interstitial sites corresponding with that network for a Markovian diffusion process.}
\centering
\begin{tabular}{ccc}
\hline
\hline
network &\multicolumn{1}{c}{$a^{-2}\cdot D_{\text{b}}$} &\multicolumn{1}{c}{$c^{-2}\cdot D_{c}$}\\
\hline
o, dh &\hspace{1 cm}\(\displaystyle2\pa{o} \frac{3\rate{o}{o(b)}}{2}+2\pa{o}\frac{3\rate{o}{dh}\rate{dh}{dh}}{2\rate{dh}{o}+3\rate{dh}{dh}}\)&\hspace{1 cm}\(\displaystyle2\pa{o}\frac{\rate{o}{o($c$)}}{4}+2\pa{o}\frac{3\rate{o}{dh}}{8}\)\\[0.3 cm]
o, h, dh, c &\hspace{1 cm}\(\displaystyle2\pa{o}\frac{3\rate{o}{c}}{4}+2\pa{h}\frac{\rate{h}{c}}{4}+2\pa{o}\frac{\rate{o}{dh}\rate{dh}{h}}{2\rate{dh}{o}+\rate{dh}{h}}\)&\hspace{1 cm}\(\displaystyle2\pa{o}\frac{\rate{o}{o($c$)}}{4}+2\pa{o}\frac{3\rate{o}{dh}}{8}+2\pa{h}\frac{3\rate{h}{c}}{8}\)\\[0.3 cm]
o, h, c &\hspace{1 cm}\(\displaystyle2\pa{o}\frac{3\rate{o}{c}}{4}+2\pa{h}\frac{\rate{h}{c}}{4}+2\pa{o}\rate{o}{h}\)&\hspace{1 cm}\(\displaystyle2\pa{o}\frac{\rate{o}{o($c$)}}{4}+2\pa{o}\frac{3\rate{o}{h}}{8}+2\pa{h}\frac{3\rate{h}{c}}{8}\)\\[0.3 cm]
t, o &\hspace{1 cm}\(\displaystyle 4\pa{t}\frac{\rate{t}{o}}{2}\)&\hspace{1 cm}\(\displaystyle2\pa{o}\frac{\rate{o}{o($c$)}}{4}+4\pa{t}\frac{3\rate{t}{o}\rate{t}{t}}{24\rate{t}{o}+16\rate{t}{t}}\)\\[3pt]
\hline
\hline
\end{tabular}
\label{tab:Expression}
\end{table}

Table~\ref{tab:Expression} lists analytical expressions for diffusivity based on the active diffusion pathways formed by the stable sites, in terms of occupation probabilities and transition rates.  We follow the approach of near-equilibrium thermodynamics to calculate the diffusivity $D$ by finding a steady state solution for the system in equilibrium distribution with a small perturbation in the chemical potential gradient of the solute\cite{TrinkleElastodiffusivity2016}. The derived analytical expressions for solute diffusivity are made up of bare mobilities and correlation effects. Table~\ref{tab:Expression} lists the term-by-term contributions to the basal diffusivity $D_{\text{b}}$ and the $c$-axis diffusivity $D_{c}$ from each type of transition. The bare mobility terms have the form of a site probability multiplied by a transition rate. The correlation effects are present in dh-o, dh-dh and dh-h transitions which contribute to the basal diffusivity as well as in t-o and t-t transitions which contribute to the $c$-axis diffusivity. Each of these networks show correlation as the jumps from particular sites (dh and t) are \textit{unbalanced}: the sum $\sum_\beta \rate{$\alpha$}{$\beta$}\mathbf{\delta x}_{\alpha-\beta} \ne 0$ for displacements $\mathbf{\delta x}_{\alpha-\beta}$ from site $\alpha$ to $\beta$. This leads to a correlated random walk where, for example, if an interstitial is in a tetrahedral site with a low t-t barrier it is very likely to be in that same tetrahedral site after two jumps; hence, a large (anti)correlation between the displacement vector in subsequent jumps. The analytical expressions are applicable in any hcp crystal for any solute having a set of interstitial sites corresponding with that network for a Markovian diffusion process. Our expression for the set of sites \{o, h, c\} agree with the expression for O diffusing in $\alpha$-Ti\cite{Wu2011}. In the case of t-t jumps which tend to have low barriers, the assumption of ``independent'' tetrahedral sites becomes invalid; instead, the pair is similar to a superbasin which thermalizes rapidly, and the $\rate{t}{t}$ disappears from the diffusivity as $\rate{t}{t}\to\infty$. The site energies and site prefactors, as well as the attempt frequencies and transition state energies of all the transitions for B, C, N and O, is available in tabular form\cite{ComputationalData}.

\begin{figure}
\centering
\includegraphics[width=0.5\textwidth]{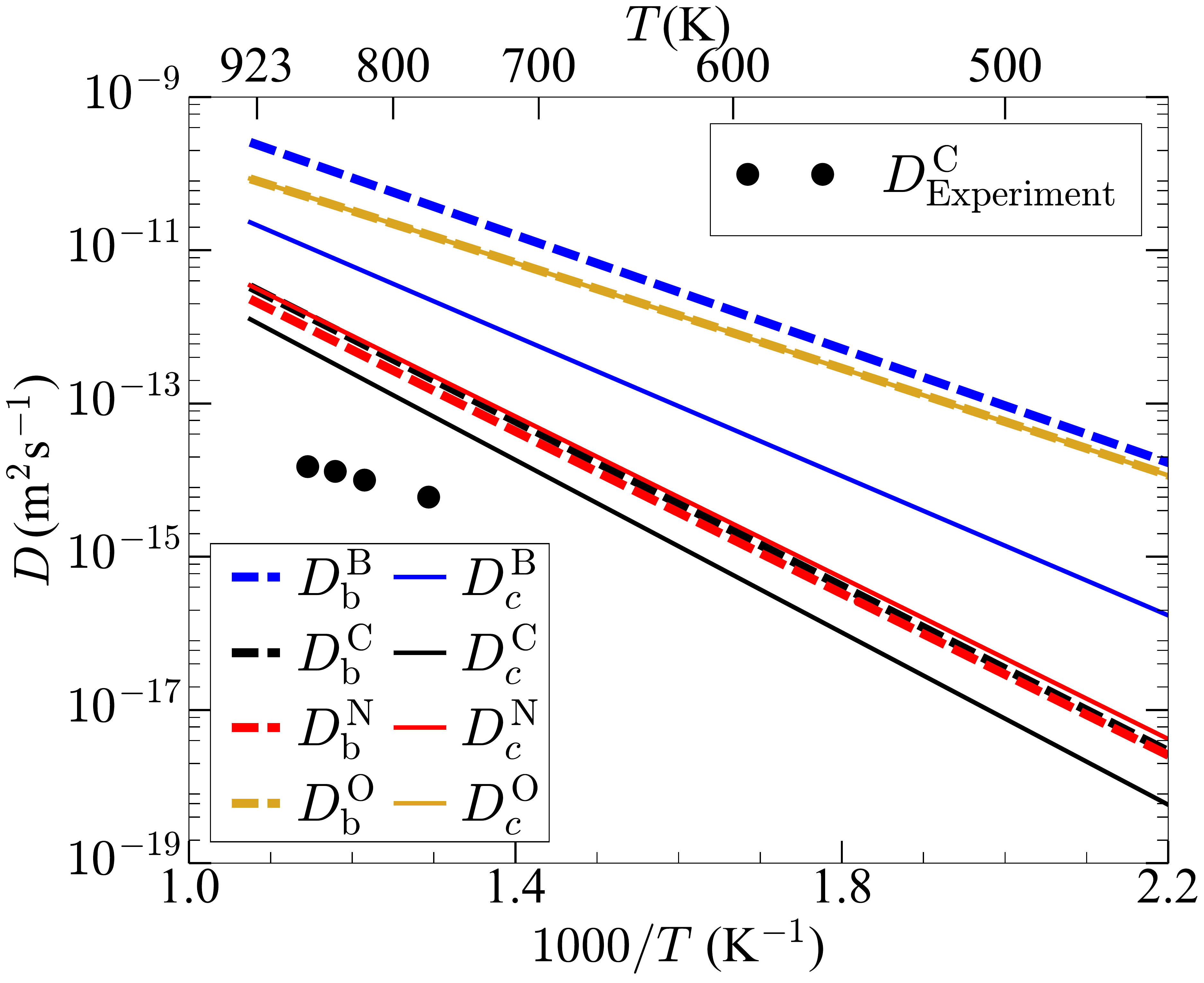}
\caption{(color online) Analytical results for diffusivities in the basal plane ($D_{\text{b}}^{\text{X}}$) and along the $c$-axis ($D_{c}^{\text{X}}$) of Mg for interstitial solute X = B, C, N and O. Diffusion of O is isotropic while diffusion of B and C is slower along the $c$-axis than in the basal plane and diffusion of N is faster along the $c$-axis than in the basal plane. The analytical expressions listed in Table~\ref{tab:Expression} are employed to compute the variation of diffusivity with temperature. Also shown is the diffusivity of C ($D_{\text{Experiment}}^{\text{C}}$), determined experimentally by Zotov $et. al$\cite{Zotov1976} at four temperatures between 773--873K.}
\label{fig:Diffusionplot}
\end{figure}

Figure~\ref{fig:Diffusionplot} shows that O diffuses isotropically while B, C, and N diffuse anisotropically. B and C diffuse faster in the basal plane than along the $c$-axis  while N diffuses faster along the $c$-axis than in the basal plane. The analytical expressions in Table~\ref{tab:Expression} give the diffusivity as a function of temperature. For all temperatures from 300K to 923K (the melting point of Mg), the basal diffusivities of the four solutes follow $D_{\text{b}}^{\text{B}}>D_{\text{b}}^{\text{O}}>D_{\text{b}}^{\text{N}} \approx D_{\text{b}}^{\text{C}}$ and the $c$-axis diffusivities follow $D_{c}^{\text{O}}>D_{c}^{\text{B}}>D_{c}^{\text{N}} > D_{c}^{\text{C}}$. Zotov $et. al$\cite{Zotov1976} measured the diffusivity of C experimentally in the temperature range of 773--873K (500--600$^\circ\text{C}$) and our results overestimate their measured diffusivity by a factor of 10--80. With only the single experiment for comparison, it is difficult to assess the source of the discrepancy.

\begin{table}
\caption{The Arrhenius fitting parameters for basal ($D_{\text{b}}^{\text{X}}$) and  $c$-axis ($D_{c}^{\text{X}}$) diffusivities through active networks of sites for interstitial solute X = B, C, N, and O. The diffusivities vary with temperature according to the Arrhenius model  $D=D_{0}\cdot\exp(-Q/\kT)$, where $D_{0}$ is the diffusivity prefactor, $Q$ is the activation energy of diffusion, \textit{T} is temperature in K, and $\kB$ is the Boltzmann constant. The comparison of energy barriers from Fig. \ref{fig:Barrier} to the activation energy $Q$ gives the dominant transition.}
\begin{tabular}{cccccc}
\hline
Solute	&Network 		&\multicolumn{2}{c}{$D_{\text{b}}^{\text{X}}$} 		&\multicolumn{2}{c}{$D_{c}^{\text{X}}$}\\
       		X&&$D_{0}\ (\mathrm{m^2s^{-1}})$  	&$Q\ (\mathrm{eV})$ 		&$D_{0}\ (\mathrm{m^2s^{-1}})$	&$Q\ (\mathrm{eV})$ \\
\hline
B		&o, dh	&2.52\e{-6}							&0.74						&1.83\e{-6}						&0.90				\\
C		&o, h, dh, c	&2.07\e{-6}							&1.07						&1.38\e{-6}						&1.11				\\
N		&o, h, c	&1.42\e{-6}							&1.05						&1.58\e{-6}						&1.04				\\
O		&o, t	&0.49\e{-6}							&0.69						&0.52\e{-6}						&0.69				\\
\hline
\end{tabular}
\label{tab:DiffusivityTable}
\end{table}

Table~\ref{tab:DiffusivityTable} lists the activation energies and diffusivity prefactors obtained from Arrhenius fits to the diffusivity plots (Fig.~\ref{fig:Diffusionplot}). For each solute, the comparison between the activation energy for diffusion $Q$ and the migration energies of individual transitions (see Fig.~\ref{fig:Barrier}) indicates the dominant type of transition that contributes most to diffusion. In the case of O, the migration energy of t-o transition is 0.70 eV which is close to the activation energy of 0.69 eV, so this transition contributes more than the other transitions to both diffusivities. Similarly, o-o basal and o-c transitions dominate for basal diffusion of B and C, respectively, while o-dh transitions dominate for $c$-axis diffusion of both these solutes. However, for N, all transitions except o-o along $c$ axis have similar energies, so it is likely that more than one transition type contributes to both diffusivities. 

\section{Elastic dipole tensor}

The elastic dipole tensor quantifies the elastic interaction energy between an external strain field and the point defect in the small strain limit. The dipole tensor is equal to the negative derivative of elastic energy $E$ with respect to strain $\underline{\varepsilon}$. The elastic dipole components $P_{ij}$ are computed from the stress tensor $\underline{\sigma}$ after relaxing the ions while keeping the supercell shape and volume $V$ fixed in the presence of the interstitial\cite{Clouet2008},
\begin{equation}
P_{ij}=-\frac{dE}{d\varepsilon_{ij}} \approx \sigma_{ij}V.
\end{equation}

The elastic dipole tensor determines the change in site energies and transition state energies of interstitial solutes due to small strain. The site energy $E_{\alpha(\textbf{s})}(\underline{\varepsilon})$ of $\alpha$ with orientation vector \textbf{s} under small strain $\underline{\varepsilon}$ is approximated by the linear relation
\begin{equation}
E_{\alpha(\textbf{s})}(\underline{\varepsilon})\approx E_{\alpha}(0)-\sum_{ij}P_{\alpha(\textbf{s}),ij}\varepsilon_{ij},
\label{eq:strainsiteenergy}
\end{equation}
where $E_{\alpha}(0)$ is the site energy of $\alpha$ in the unstrained cell and $P_{\alpha(\textbf{s}),ij}$ are the elastic dipole components of site $\alpha$ with orientation \textbf{s}. In the infinitesimal strain limit, the sites and network topology remains unchanged; with larger finite strains, sites may become unstable or change the network topology, which requires a new analysis of network. The vector \textbf{s} distinguishes the multiple sites of the same type which are present in an hcp unit cell. The orientation of c site is defined as the vector connecting it to the nearest o site and the orientation of dh site is defined as the vector connecting it to the nearest h site. In a hcp unit cell (see Fig. \ref{fig:Unitcell}), there are two o, two h, four t, six c and six dh sites. In an unstrained cell, multiple sites of the same type have the same energy. However, strain can cause these sites to become nonequivalent in energy depending on their elastic dipole tensor which may depend on their site orientation. The  dipoles for o, h, and t sites are independent of their orientation vector while the dipole for c and dh sites depend on their orientation vector. Similarly, the transition state energy $E_{\alpha(\textbf{s})\textnormal{-}\beta(\textbf{s}')}^{\textbf{v}}(\underline{\varepsilon})$ for site $\alpha$ of orientation \textbf{s} to site $\beta$ of orientation $\textbf{s}'$ under strain is  
\begin{equation}
E_{\alpha(\textbf{s})\textnormal{-}\beta(\textbf{s}')}^{\textbf{v}}(\underline{\varepsilon}) \approx E_{\alpha\textnormal{-}\beta}(0)-\sum_{ij}P_{\alpha(\textbf{s})\textnormal{-}\beta(\textbf{s}') ,ij}^{\textbf{v}}\varepsilon_{ij},
\label{eq:straintransitionenergy}
\end{equation}
where \textbf{v} is the vector from site $\alpha$ to $\beta$, $E_{\alpha\textnormal{-}\beta}(0)$ is the \textbf{v}-independent transition state energy in the unstrained cell and $P_{\alpha(\textbf{s})\textnormal{-}\beta(\textbf{s}') ,ij}^{\textbf{v}}$ are the elastic dipole components at the transition state corresponding to \textbf{v}. As discussed previously in Fig.~\ref{fig:Network}, there are multiple transitions of the same type distinguished through their transition vectors \textbf{v}. In a strained cell, these transitions can have different transition state energies depending on their dipole tensors which may depend on their transition vectors. 

\begin{table}
\caption{Elastic dipole tensors $\underline{P}$ at representative interstitial sites for B, C, N, and O in Mg. The symmetric elastic dipole tensor is diagonal along three principal axes $\mathbf{e_{1}}$, $\mathbf{e_{2}}$, and $\mathbf{e_{3}}$ and has units of eV. For c and dh sites, the dipole tensors and their axes depend on the orientations \textbf{s} of the sites with respect to the nearest o and h sites, respectively, whereas the dipole tensors for o, t and h sites are independent of orientation. The possible orientations of dh sites with respect to an h site are $[\overline{1}100]$, $[10\overline{1}0]$ and [$0\overline{1}10]$, and the orientations of c sites with respect to an o site are $[2\overline{1}\overline{1}0]$, $[\overline{1}\overline{1}20]$ and $[\overline{1}2\overline{1}0]$. Here the dipole tensor of each type of site is given for one representative \textbf{s}, and other tensors with different \textbf{s} are obtained by applying the appropriate point group operations on the representative dipole tensor. }
\begin{tabular}{ccc D{,}{\pm}{0} D{,}{\pm}{0} D{,}{\pm}{0} ccc}
\hline
Solute		&Site	&Orientation (\textbf{s})		&P_{11}		&P_{22}		&P_{33}		&$\mathbf{e_{1}}$		&$\mathbf{e_{2}}$		&$\mathbf{e_{3}}$  \\
\hline
B			&o		& any	&$2.38$			&$2.38$			&$2.55$			&\multicolumn{2}{c}{orthogonal basal vectors}&[$0001$]	\\
			&dh &[$\overline{1}100$]		&$11.03$			&$0.04$			&-0.49		&[$11\overline{2}0$]&[$\overline{1}100$]&[$0001$]	\\
C			&o	&any		&$1.08$			&$1.08$			&$0.24$			&\multicolumn{2}{c}{orthogonal basal vectors}&[$0001$]	\\
			&h		&any	&$4.74$			&$4.74$			&-1.10		&\multicolumn{2}{c}{orthogonal basal vectors}&[$0001$]	\\
			&c &[$2\overline{1}\overline{1}0$]			&$6.59$	        &$4.20$          &-5.18        &[$0\frac{1}{3}\overline{\frac{1}{3}}\frac{1}{2}$]&[$2\overline{1}\overline{1}0$]&[$01\overline{1}\overline{\frac{3}{4}}$]	\\
			&dh &[$\overline{1}100$]			&$8.94$			&-0.22		&-0.86		&[$11\overline{2}0$]&[$\overline{1}100$]&[$0001$]		\\
N			&o		&any	&$0.00$			&$0.00$			&-1.39		&\multicolumn{2}{c}{orthogonal basal vectors}&[$0001$]			\\
			&h		&any	&$3.22$			&$3.22$			&-1.81		&\multicolumn{2}{c}{orthogonal basal vectors}&[$0001$]				\\
			&c &[$2\overline{1}\overline{1}0$]			&$4.22$        &$4.31$ &-5.39           &[$0\frac{1}{3}\overline{\frac{1}{3}}\frac{1}{2}$]&[$2\overline{1}\overline{1}0$]&[$01\overline{1}\overline{\frac{3}{4}}$]	\\      
O			&o	&any		&-0.15		&-0.15		&-1.76		&\multicolumn{2}{c}{orthogonal basal vectors}&[$0001$]				\\
			&t		&any	&$2.06$			&$2.06$			&$0.79$ 		&\multicolumn{2}{c}{orthogonal basal vectors}&[$0001$]				\\
\hline
\end{tabular}
\label{tab:Sitedipole}
\end{table}  

\setlength{\tabcolsep}{1.5pt}
\begin{table}
\caption{Elastic dipole tensors $\underline{P}$ at representative transition states for B, C, N, and O in Mg. The transition state from site $\alpha$ to site $\beta$ is denoted by $\alpha\textnormal{-}\beta$, and \textbf{v} is the vector connecting these two sites. The symmetric elastic dipole tensor is diagonal along three principal axes  $\mathbf{e_{1}}$, $\mathbf{e_{2}}$, and $\mathbf{e_{3}}$ and has units of eV. The dipole tensor of an equivalent transition with a different \textbf{v} is obtained by applying the appropriate point group operation to the given dipole tensor. The variable $x$ for B and C is 0.197 and 0.238,  respectively, and variable $z$ for O is 0.153. The values of $x$ and $z$ are obtained from the relaxed position of dh and t sites in the Mg supercell, respectively.}
\begin{tabular}{lccD{,}{\pm}{0} D{,}{\pm}{0} D{,}{\pm}{0}ccc}
\hline
Solute		&$\alpha\textnormal{-}\beta$ &Transition (\textbf{v})	&P_{11}		&P_{22}		&P_{33}		& $\mathbf{e_{1}}$ &$\mathbf{e_{2}}$ &$\mathbf{e_{3}}$  \\
\hline
B			&o-o &$[000\frac{1}{2}]$			&5.34		&5.34			&-3.58		&\multicolumn{2}{c}{orthogonal basal vectors}&[$0001$]	\\
			&o-o &$\frac{1}{3}[2\overline{1}\overline{1}0]$			&-3.08			&2.74			&8.56			&$[0\overline{1}1\overline{\frac{3}{4}}]$&$[2\overline{1}\overline{1}0]$&[$0\overline{\frac{1}{3}}\frac{1}{3}\frac{1}{2}$]	\\
			&o-dh &[$\overline{x}x0\frac{1}{4}$]		&$10.99$			&$0.19$			&-0.61		&[$11\overline{2}0$]&[$1\overline{1}0\frac{1}{2}$]&[$\overline{1}10\frac{9}{4}$]	\\

			&dh-dh &$(x-\frac{1}{3})[2\overline{1}\overline{1}0]$		&7.69			&4.25			&-0.20		&[$01\overline{1}0$]&[$2\overline{1}\overline{1}0$]&[$0001$]	\\
C			&o-o &$[000\frac{1}{2}]$			&3.63			&3.63			&-3.22		&\multicolumn{2}{c}{orthogonal basal vectors}&[$0001$]	\\
			&o-c &$\frac{1}{6}[2\overline{1}\overline{1}0]$]			&-3.93	        &1.57         &5.19         &$[0\overline{1}1\overline{\frac{3}{4}}]$&[$2\overline{1}\overline{1}0$]&[$0\overline{\frac{1}{3}}\frac{1}{3}\frac{1}{2}$]	\\
			&o-dh &[$\overline{x}x0\frac{1}{4}$] &-2.60 &0.58 &7.28 &$[\overline{1}10\sqrt{\frac{3}{8}}]$ &$[1\overline{1}0\sqrt{\frac{27}{8}}]$ &$[11\overline{2}0]$\\
			&h-c &$[0\frac{1}{6}\overline{\frac{1}{6}}\frac{1}{4}]$ &3.02 &-0.68 &4.45 &$[2\overline{1}\overline{1}0]$ &$[0\frac{1}{6}\overline{\frac{1}{6}}\frac{1}{4}]$&$[01\overline{1}\overline{\frac{3}{4}}]$\\	
			&h-dh &$(x-\frac{1}{3})[\overline{1}100]$ &7.59&1.07&-1.01&$[11\overline{2}0]$ &$[\overline{1}100]$ &$[0001]$\\	
N			&o-o &$[000\frac{1}{2}]$			&2.58			&2.58			&-1.16		&\multicolumn{2}{c}{orthogonal basal vectors}&[$0001$]	\\
			&o-c &$\frac{1}{6}[2\overline{1}\overline{1}0]$]			&-4.08	        &3.23         &3.58         &$[0\overline{1}1\overline{\frac{3}{4}}]$&[$2\overline{1}\overline{1}0$]&[$0\overline{\frac{1}{3}}\frac{1}{3}\frac{1}{2}$]	\\
			&o-h &[$\overline{\frac{1}{3}}\frac{1}{3}0\frac{1}{4}$] &-2.08&0.31 &4.61 &$[0.57,  \overline{0.57},0,\overline{0.10}]$ &$[0.09, \overline{0.09},0,0.60]$  &$[11\overline{2}0]$\\
			&h-c &$[0\frac{1}{6}\overline{\frac{1}{6}}\frac{1}{4}]$ &3.72 &-4.14 &3.94 &$[2\overline{1}\overline{1}0]$ &$[0\overline{1}1\overline{\frac{3}{4}}]$&$[0\overline{\frac{1}{3}}\frac{1}{3}\frac{1}{2}]$\\
O			&o-o &$[000\frac{1}{2}]$			&2.37			&2.37			&1.76		&\multicolumn{2}{c}{orthogonal basal vectors}&[$0001$]	\\
&t-t &$[000(\frac{1}{2}-2z)]$			&1.97			&1.97			&-1.67		&\multicolumn{2}{c}{orthogonal basal vectors}&[$0001$]	\\
&o-t &$[\overline{\frac{1}{3}}\frac{1}{3}0z]$			&0.67			&1.37			&2.07		&$[0.12, \overline{0.12},0,\overline{0.60}]$ &$[0.56, \overline{0.56},0,0.12]$  &$[11\overline{2}0]$\\	\\				
\hline
\end{tabular}
\label{tab:Transitiondipole}
\end{table}

Tables~\ref{tab:Sitedipole} and~\ref{tab:Transitiondipole} list the components of the elastic dipole tensor at representative interstitial sites with orientations \textbf{s}, and representative transition states with transition vectors \textbf{v}. We diagonalize the elastic dipole tensors along three principal axes ($\mathbf{e_{1}}$, $\mathbf{e_{2}}$, $\mathbf{e_{3}}$), and report the diagonalized entries entries ($P_{11}, P_{22}, P_{33}$) and principal axes. From Table~\ref{tab:Sitedipole}, the elastic dipole components in the two orthogonal basal directions are equal for o, h, and t sites due to the basal symmetry of these sites.        The trace of the elastic dipole for N and O at o sites is negative, leading to the volumetric contraction upon cell relaxation, in contrast to the other interstitial sites. The ground state configuration of N undergoes volume contraction on cell relaxation while the ground state configuration of B, C, and O undergoes volume expansion on cell relaxation. In the case of the dh site, its two nearest Mg atoms experience larger atomic forces compared to other Mg atoms therefore, the elastic dipole for the dh site has the largest component in the [$11\overline{2}0$] direction which connects these two nearest Mg atoms. From Table~\ref{tab:Transitiondipole}, most of the transition states break the symmetry of the crystal except for the o-o and t-t transitions along the $c$-axis which obey the basal symmetry. Because of the basal symmetry, the transition state energies of the o-o ($c$-axis) and t-t transitions with different \textbf{v} remain equivalent in the strained cell while the same is not true for the other types of transitions.

Elastic dipole tensors for symmetry-equivalent sites with different \textbf{s}, and symmetry-equivalent transitions with different \textbf{v}, are obtained by point group operations on the representative dipole tensors in Tables~\ref{tab:Sitedipole} and~\ref{tab:Transitiondipole}. For example, the three c sites in the basal plane with different orientations ($[2\overline{1}\overline{1}0]$, $[\overline{1}\overline{1}20]$ and $[\overline{1}2\overline{1}0]$) are all related Wyckoff sites, that are transformed by $120^\circ$ rotations about the $c$-axis; call that transformation matrix $R$. The dipole tensors for the other two equivalent sites $\mathbf{s}'$ are
\begin{equation}
\underline{P}_{\alpha(\textbf{s}')}=R \underline{P}_{\alpha(\textbf{s})}R^{T}
\label{eq:dipoletransformation}
\end{equation}
where $\underline{P}_{\alpha(\textbf{s})}$ is the representative dipole tensor and $R$ transforms $\textbf{s}$ to $\textbf{s}'$. Similarly, the dipole tensors of all the other sites are calculated using their associated transformation matrices. The same operations are carried out for all the transition state dipole tensors based on the symmetry of the transition vectors \textbf{v}. The dipole data in Cartesian basis for all these equivalent sites and equivalent transitions for B, C, N and O are available in tabular form\cite{ComputationalData}. This dipole tensor data is used to estimate changes in site energies and the changes in migration barriers of transitions under strain using Eqs.~\ref{eq:strainsiteenergy} and~\ref{eq:straintransitionenergy}, which are inputs to the elastodiffusion tensor calculations.

\section{Elastodiffusion tensor}
Strain affects the diffusivity of solutes by changing the jump vectors and migration barriers of the diffusion network. The first order strain dependence of diffusivity is represented with the elastodiffusion tensor\cite{Dederichs1978,Savino1987,Woo2000,TrinkleElastodiffusivity2016}$\underline{d}$
\begin{equation}
d_{ijkl}=\frac{\partial D_{ij}}{\partial \varepsilon_{kl}},
\end{equation}
and is derived using perturbation theory\cite{TrinkleElastodiffusivity2016,OnsagerCalc}.
The contribution $\underline{d}^{\rm{geom}}$ to the elastodiffusion tensor from the changes in jump vectors is\cite{TrinkleElastodiffusivity2016}
\begin{equation}
d_{ijkl}^{\rm{geom}}=\frac{1}{2}(D_{jk}(0)\delta_{il}+D_{il}(0)\delta_{jk}+D_{ik}(0)\delta_{jl}+D_{jl}(0)\delta_{ik}),
\end{equation} 
where $\delta_{ij}$ are the Kronecker deltas.
Hence, if the diffusivity has Arrhenius temperature dependence, then so does the geometric term in the elastodiffusion tensor.
The contribution $\underline{d}^{\rm{mb}}$ from changes in the migration barriers is determined by the elastic dipole tensors of the migration barriers and sites. The elastic dipole tensor of a transition state relative to initial site determines the rate of that transition under strain and the elastic dipole tensor of interstitial site determine the occupation probability of that site under strain. The term $\underline{d}^{\rm{mb}}$ is the sum of contributions from each transition; these contributions are proportional to the product of the inverse temperature, transition rate, and difference of transition state dipole and thermal average dipole of interstitial sites. The contribution from one transition can be represented as
\begin{equation}
  \label{eq:elastodiffusion_migrationbarrier}
  \frac{d_0}{\kT} \cdot\exp(-E/\kT)
\end{equation}
where the elastic dipole terms are absorbed in the ``prefactor'' $d_{0}$, which has units of $\text{eV}\cdot\text{m}^2\text{s}^{-1}$, and $E$ is the barrier of the dominant transition. 




The symmetry of the hexagonal closed-packed crystal reduce the number of unique elastodiffusion components to six. We use Voigt notation, similar to elastic constants, to represent the indices of the fourth rank tensor as both diffusivity and strain are symmetric second rank tensors. The reduction by symmetry is the same as the elastic constants, except that $d_{ij}$ is not necessarily equal to $d_{ji}$. In the case of hcp, the non-zero elastodiffusion elements are $d_{11} = d_{22}$, $d_{33}$, $d_{12}$, $d_{13}=d_{23}$, $d_{31}=d_{32}$, $d_{44}=d_{55}$, and $d_{66}=(d_{11}-d_{12})/2$. The change in jump vectors contributes only to $d_{11}$, $d_{33}$, $d_{44}$, and $d_{66}$. Unlike the contribution from the change in jump vectors, the change in migration barrier can contribute to all six independent components of elastodiffusion tensor and need not only be positive. 

\begin{table}
  \caption{The fitting parameters $d_{0}$ and $E$ in Eqn.~\ref{eq:elastodiffusion_migrationbarrier} for the components of the B, C, N, and O elastodiffusion tensor in Mg over 300--923K. The elastodiffusion tensor in Voigt notation has six unique components in an hcp crystal, where $d_{66}=(d_{11}-d_{12})/2$. A subset of components change sign with temperature; their transition temperature is listed in lieu of fitting parameters (c.f., Fig.~\ref{fig:Elastodiffusionplot} for the temperature dependence). The ``activation barrier'' $E$ corresponds closely to the migration barrier of the dominant transition. The $d_{12}$ and $d_{33}$ components for B, all diagonal components for C, and $d_{44}$ and $d_{66}$ for N and O are negative throughout the temperature range (i.e. have negative $d_{0}$). The negative $d_{0}$ implies that the increase in diffusivity caused by lowered migration barriers is greater than the decrease in diffusivity due to reduced jump vectors under compressive strains. For $d_{44}$ of B, the geometric contribution is dominant and is best described with an Arrhenius fit of $1.3\times 10^{-6}\text{m}^2/\text{s}\cdot\exp(-0.74/\kT)$.}
\centering
\begin{tabular}{c D{,}{\pm}{0} c D{,}{\pm}{0} c D{,}{\pm}{0} c D{,}{\pm}{0} c}
\hline
&\multicolumn{2}{c}{B}&\multicolumn{2}{c}{C}&\multicolumn{2}{c}{N}&\multicolumn{2}{c}{O}\\
&d_{0}( \mathrm{eVm^2s^{-1}}) &$E(\mathrm{eV})$ &d_{0}( \mathrm{eVm^2s^{-1}}) &$E(\mathrm{eV})$&d_{0}( \mathrm{eVm^2s^{-1}}) &$E(\mathrm{eV})$&d_{0}( \mathrm{eVm^2s^{-1}}) &$E(\mathrm{eV})$\\
\hline
$d_{11}$&\multicolumn{2}{c}{(854.7K)}&-3.0\e{-8}&0.91&\multicolumn{2}{c}{(398.4K)}&\multicolumn{2}{c}{(900.9K)}	\\
$d_{12}$&-2.9\e{-6}&0.74&5.0\e{-8}&0.94&2.5\e{-6}&1.04&\multicolumn{2}{c}{(678.0K)}\\
$d_{13}$&5.6\e{-6}&0.74&2.0\e{-6}&1.05&2.9\e{-6}&1.05&\multicolumn{2}{c}{(552.5K)}\\
$d_{31}$&5.5\e{-6}&0.90&2.2\e{-6}&1.12&1.8\e{-6}&1.04&\multicolumn{2}{c}{(865.8K)}\\
$d_{33}$&-5.3\e{-6}&0.90&-3.3\e{-7}&1.10&3.9\e{-6}&1.05&\multicolumn{2}{c}{(409.8K)}\\
$d_{44}$&1.5\e{-7}&0.78$^*$&-9.7\e{-7}&1.11&-1.2\e{-6}&1.04&-1.0\e{-8}&0.65\\
$d_{66}$&1.5\e{-6}&0.74&-4.0\e{-8}&0.93&-9.3\e{-7}&1.03&-5.0\e{-8}&0.68\\


\hline
\end{tabular}
\label{tab:ElastoTable}
\end{table}


Table~\ref{tab:ElastoTable} shows that the contribution $\underline{d}^{\rm{mb}}$ dominates over the contribution $\underline{d}^{\rm{geom}}$ due to the relatively larger values of elastic dipole tensor components compared to $\kT$ (see eqn.~\ref{eq:elastodiffusion_migrationbarrier}) for all the temperatures between 300--923K. However, the contribution $\underline{d}^{\rm{geom}}$ is greater than the contribution $\underline{d}^{\rm{mb}}$ for the $d_{44}$ component for B due to larger transition rate of o-o transition in basal plane and for the $d_{11}$ component for B and O at temperatures above crossover (discussed in the later paragraph). Equation \ref{eq:elastodiffusion_migrationbarrier} is used to fit the elastodiffusion component because of the larger contribution from $\underline{d}^{\rm{mb}}$ over $\underline{d}^{\rm{geom}}$ and also due to the dominant transition for each solute. The fitting parameter $E$ in Table~\ref{tab:ElastoTable} corresponds to the migration barrier of the dominant transition. These dominant transitions under strain is same as that in the unstrained crystal, except for the basal components $d_{11}$, $d_{12}$ and $d_{66}$ for C which are now dominated by the h-dh transition. The remaining basal component $d_{13}$ of C is governed by o-c transition and the basal components ($d_{12}$, $d_{13}$ and $d_{66}$) and $d_{44}$ of B are governed by o-o(b)transition. The non-basal components ($d_{31}$ and $d_{33}$) are governed by o-dh transition for both B and C. The isotropic o-t transition is dominant for all the components for O, and in N, both o-h and h-c transitions, which have similar migration barriers, contribute to elastodiffusion components.

\begin{figure}
\centering
\includegraphics[width=0.5\textwidth]{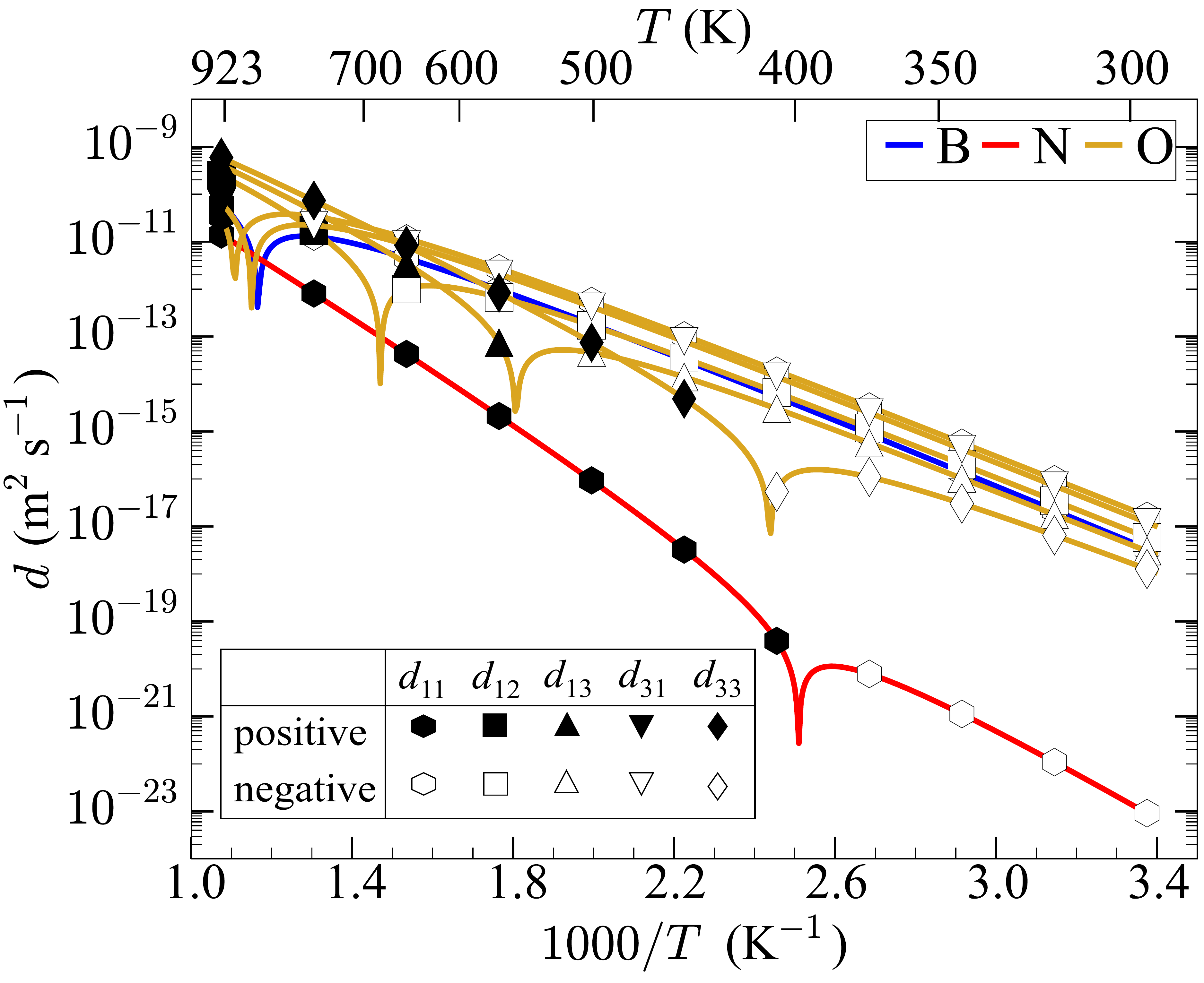}
\caption{(color online) Components of the elastodiffusion tensor $\underline{d}$ that change sign as a function of temperature, for B, N, and O. The magnitudes of each component are shown with filled symbols for positive values and unfilled for negative values. For a component, changes of sign is observed as a dip in the curve and the crossover temperatures is listed in Table~\ref{tab:ElastoTable}. The sign inversion of these components is caused by two competing mechanisms, which dominate at either low or high temperatures. Five components of the elastodiffusion tensor for O change sign, and each component has a different crossover temperature.}
\label{fig:Elastodiffusionplot}
\end{figure}

Figure~\ref{fig:Elastodiffusionplot} shows that five of the elastodiffusion components for oxygen change sign (fewer for B, and N) due to the small energy separation from the ground state and the metastable states, while for B, C, and N the energy separation is significant. The change in sign from positive (filled symbol) to negative (unfilled symbol) is observed as dips in the logarithm of the magnitude of $d$ and the associated crossover temperature is listed in parenthesis in Table~\ref{tab:ElastoTable} for these components. The sign inversion of these components is due to the competing mechanism dominating over different temperature which we observe as different slopes on opposite side of crossover. The sign inversion of $d_{12}$, $d_{13}$, $d_{31}$, and $d_{33}$ for O is due to the large variation in thermally averaged elastic dipole tensor of sites, which occurs because of the low energy separation of 0.21 eV between o and t sites. The difference between the transition state dipole and the thermally averaged dipole contributes to the elastodiffusion component sign changes with temperature as the o and t sites have different elastic dipoles. However, for $d_{11}$ for B and O, the sign inversion is due to the competition between the negative contribution of $\underline{d}^{\rm{mb}}$ and positive contribution of $\underline{d}^{\rm{geom}}$, where the former dominates below the crossover temperature (due to smaller value of $\kT$ compare to dipole tensor, c.f. Eqn.~\ref{eq:elastodiffusion_migrationbarrier}) and the latter dominates above the crossover temperature. For the component $d_{11}$ of N, sign inversion is due to the o-c transition dominating above the crossover temperature while the o-h transition dominates below the crossover. The sign inversion behavior of different components suggest that the diffusivity under strain will have contrasting features around a specific temperature which we observe for the activation volume of diffusion and for the effect of thermal expansion on diffusion.

\subsection*{Activation volume of diffusion}
The elastodiffusion tensor together with the elastic compliance tensor computes the activation volume of diffusion. The activation volume of diffusion $V_{ij}$ describes the pressure $p$ dependence of diffusivity as
\begin{equation}
D_{ij}(p)=D_{ij}(0)\cdot \text{exp}(-\frac{pV_{ij}}{\kT}),
\label{eq:pressureD}
\end{equation}
where $D_{ij}(0)$ is the diffusivity tensor components at zero pressure. The activation volume is calculated using
\begin{equation}
\begin{split}
V_{ij}&=-(D_{ij}(0))^{-1}\kT \left.\frac{\partial D_{ij}}{\partial p}\right|_{p=0}\\
 &=(D_{ij}(0))^{-1}\kT \sum_{kl} d_{ijkk}S_{kkll}
\end{split}
\label{eq:AV}
\end{equation}
where $\underline{d}$ is the elastodiffusivity tensor and $\underline{S}$ is the elastic compliance tensor. In the case of interstitial diffusion, the activation volume is equal to the migration volume of a jump: the volume change between the transition state and initial state\cite{Mehrer2011}. 

\begin{figure}
  \centering
   \includegraphics[width=0.5\textwidth]{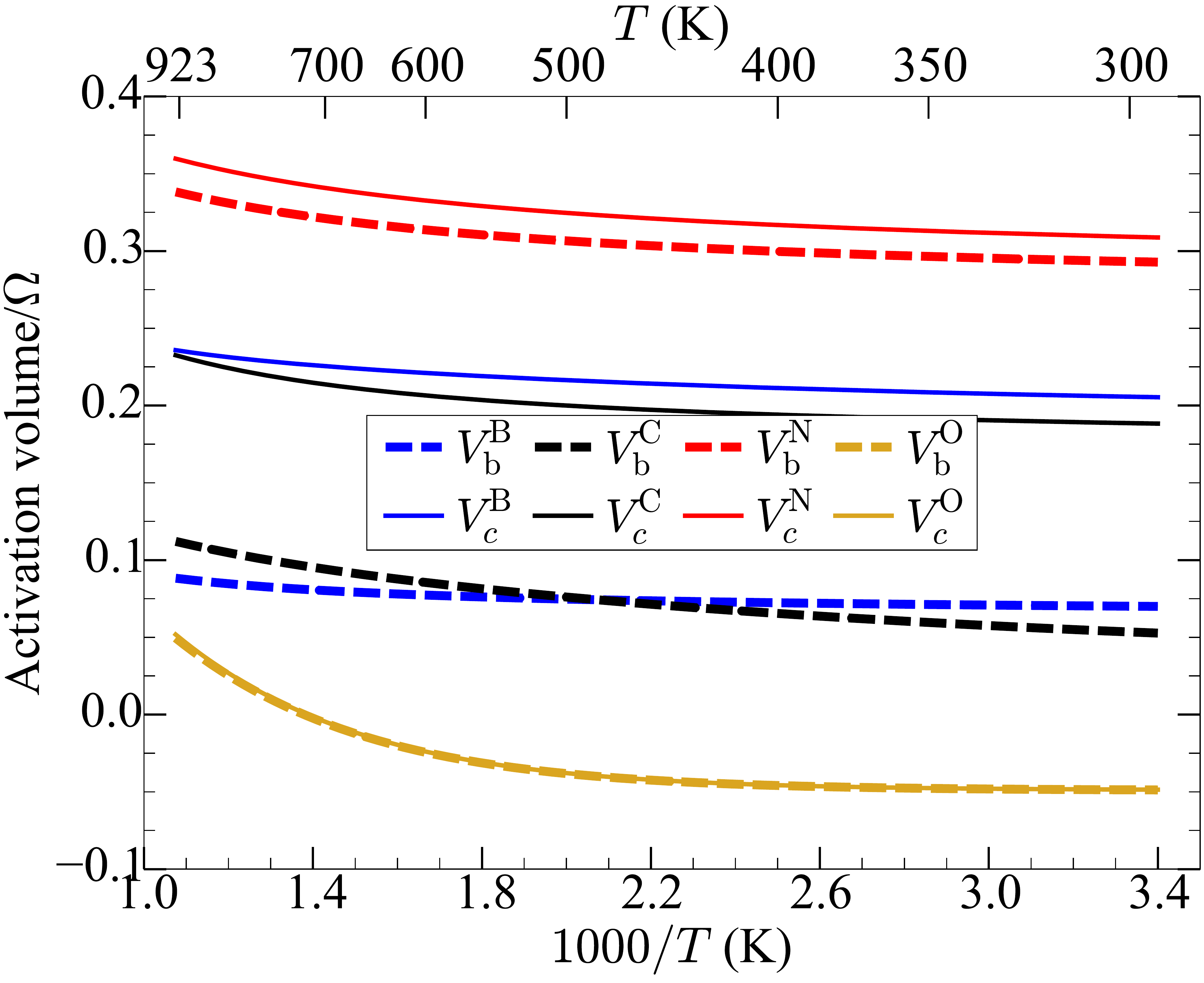}
  \caption{(color online) Activation volume for basal diffusion $V_{\text{b}}$ and $c$-axis diffusion $V_{c}$, relative to the Mg atomic volume $\Omega = 22.84$ \AA$^3$ per atom as a function of temperature for B, C, N and O. For both basal and $c$-axis diffusion, the activation volume of O is isotropic and negative below 740K while it remains positive for B, C and N. The activation volume for all the solutes increases with increasing temperatures, in part, as the elastic constants soften as temperature increases\cite{Greeff1999}. This increase is $\sim$14\% for basal activation volume and $\sim$15\% for $c$-axis activation volume for all the solutes at 923K.}
\label{fig:AV}
\end{figure}

Figure~\ref{fig:AV} shows that the activation volume for O diffusion is isotropic and negative below 740K, which leads to an increase in basal and $c$-axis diffusivities under hydrostatic pressure. The activation volumes for B, C and N diffusion remain positive throughout the temperature range, with N having the largest activation volume. For O diffusion below 740K, the dominating t-o transition has negative migration volume, while the dominating transitions for the diffusion of other solutes have positive migration volumes. Negative activation volume has also been observed experimentally for C diffusing in hcp-Co\cite{Wuttig1971} and in $\alpha$-Fe\cite{Bosman1960}, and their magnitudes are comparable to the activation volume computed for O diffusion in Mg. 
Due to the temperature-induced softening of the elastic constants\cite{Greeff1999}, the activation volume of basal and $c$-axis diffusion increases by $\sim$14\% and $\sim$15\% from 300K to 923K for all four solutes. 
\begin{figure}[htp]
\centering
\includegraphics[width=0.5\textwidth]{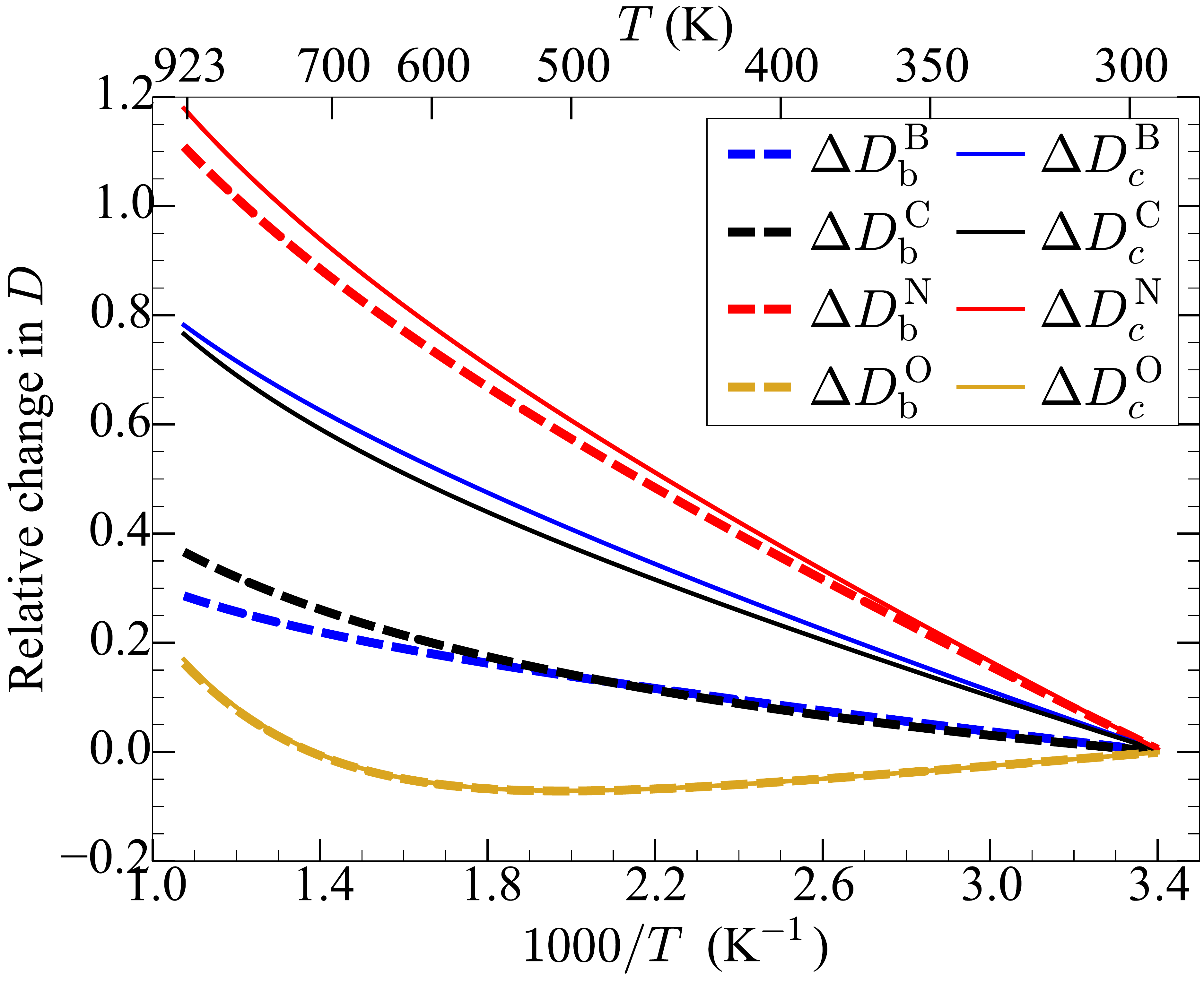}
\caption{(color online)
Change in basal and $c$-axis diffusivity due to thermal strain, relative to the strain free diffusivity for B, C, N and O. The thermal strain is nearly isotropic and linear over the entire temperature range, to a maximum value of 2\%\ at the melting temperature of 923K. The effect of thermal expansion is largest for N, for which the diffusivity doubles approaching melting, and smallest for O. Below 740K, O diffusivity decreases relative to its strain free diffusivity---due to the negative activation volume---unlike the other three solutes.}
\label{fig:ThermalD}
\end{figure}

\subsection*{Thermal expansion effect on diffusion}
Figure~\ref{fig:ThermalD} shows that thermal expansion increases the diffusivity of B, C and N, but decreases the diffusivity of O up to 740K. The fit of experimental thermal expansion data to temperature\cite{TEdata} is used to estimate thermal strain. Thermal expansion is nearly isotropic in the temperature range 300K to 923K, reaching a maximum value of 2\%. For B, C and N, both basal and $c$-axis diffusivities increase upon thermal expansion, with N experiencing the largest effect of more than 100\% increase in diffusivity at $T>$ 816K. Under thermal strain, O diffusion remains isotropic due to the dominating t-o transitions which contribute equally to diffusion in the basal plane and along the $c$-axis. Above 740K the O diffusivity is greater compared to its strain free diffusivity as expected due to thermal expansion. However, below 740K the O diffusivity is lower compared to its strain free diffusivity. This non-montonic behavior of O diffusivity with thermal expansion is due to the sign inversion of five of the elastodiffusion tensor components.

\section{Conclusion}
We determine the stable interstitial sites, migration barriers, diffusivities, and elastodiffusion tensors for B, C, N and O in Mg. We find a new stable distorted hexahedral site that B and C can occupy in Mg. Analytical expressions for interstitial diffusion in bulk hcp crystals are derived for the networks of interstitial sites. Diffusion of O is isotropic due to dominating isotropic t-o transitions while B and C have faster basal diffusion compared to $c$-axis diffusion and N have slower basal diffusion compared to $c$-axis diffusion. This shows that diffusion depends on the diffusion network formed by sites and their energetics, which varies from solute to solute. The elastodiffusion tensor captures the effect of strain on diffusivity by summing the contributions from changes in jump vectors and changes in migration barriers. For B, C, N and O in Mg, the contribution to elastodiffusion components due to changes in migration barriers dominates over the contribution from changes in jump vectors with a few exceptions. There are a few elastodiffusion components which change sign at crossover temperature due to competing mechanisms. In the case of O, five of the elastodiffusion components change sign, which leads to negative activation volume below 740K and decreased diffusivity upon thermal expansion. This behavior of O as an interstitial defect is counterintuitive because interstitial diffusivity is expected to decrease under compression as transition states are usually ``smaller.'' We see that N in its ground state (octahedral) contracts the crystal upon relaxation while it has the positive activation volume; O in its ground state (tetrahedral) expands the crystal on relaxation while having a negative activation volume. This shows that elastic dipole tensor of transition states plays a vital role along with the energetics of sites. Our study of interstitial solute diffusion under strain can be extended for other crystal structures and interstitial defects. Finally, understanding interstitial solute kinetics under strain can be helpful in studying the solute diffusivity in the heterogeneous strain fields due to dislocations or other defects.

\begin{acknowledgments}
Figures~\ref{fig:Unitcell} and~\ref{fig:Network} are generated using the Jmol package\cite{Jmol}. The authors thank Graeme Henkelman for helpful conversations. This research was supported by the U.S. Office of Naval Research under the grant N000141210752 and the National Science Foundation Award 1411106, with computing resources provided by the University of Illinois campus cluster program. The full tabular data is archived by NIST at \url{materialsdata.nist.gov}; see Ref.~\onlinecite{ComputationalData}. The authors also thank the Library Service at Los Alamos National Laboratory for locating a copy of Ref.~\onlinecite{Zotov1976}, and Yulia Maximenko at Univ. Illinois, Urbana-Champaign, Dept. of Physics for her translation help.
\end{acknowledgments}


\begin{thebibliography}{34}%
\makeatletter
\providecommand \@ifxundefined [1]{%
 \@ifx{#1\undefined}
}%
\providecommand \@ifnum [1]{%
 \ifnum #1\expandafter \@firstoftwo
 \else \expandafter \@secondoftwo
 \fi
}%
\providecommand \@ifx [1]{%
 \ifx #1\expandafter \@firstoftwo
 \else \expandafter \@secondoftwo
 \fi
}%
\providecommand \natexlab [1]{#1}%
\providecommand \enquote  [1]{``#1''}%
\providecommand \bibnamefont  [1]{#1}%
\providecommand \bibfnamefont [1]{#1}%
\providecommand \citenamefont [1]{#1}%
\providecommand \href@noop [0]{\@secondoftwo}%
\providecommand \href [0]{\begingroup \@sanitize@url \@href}%
\providecommand \@href[1]{\@@startlink{#1}\@@href}%
\providecommand \@@href[1]{\endgroup#1\@@endlink}%
\providecommand \@sanitize@url [0]{\catcode `\\12\catcode `\$12\catcode
  `\&12\catcode `\#12\catcode `\^12\catcode `\_12\catcode `\%12\relax}%
\providecommand \@@startlink[1]{}%
\providecommand \@@endlink[0]{}%
\providecommand \url  [0]{\begingroup\@sanitize@url \@url }%
\providecommand \@url [1]{\endgroup\@href {#1}{\urlprefix }}%
\providecommand \urlprefix  [0]{URL }%
\providecommand \Eprint [0]{\href }%
\providecommand \doibase [0]{http://dx.doi.org/}%
\providecommand \selectlanguage [0]{\@gobble}%
\providecommand \bibinfo  [0]{\@secondoftwo}%
\providecommand \bibfield  [0]{\@secondoftwo}%
\providecommand \translation [1]{[#1]}%
\providecommand \BibitemOpen [0]{}%
\providecommand \bibitemStop [0]{}%
\providecommand \bibitemNoStop [0]{.\EOS\space}%
\providecommand \EOS [0]{\spacefactor3000\relax}%
\providecommand \BibitemShut  [1]{\csname bibitem#1\endcsname}%
\let\auto@bib@innerbib\@empty
\bibitem [{\citenamefont {Pollock}(2010)}]{Pollock2010}%
  \BibitemOpen
  \bibfield  {author} {\bibinfo {author} {\bibfnamefont {T.~M.}\ \bibnamefont
  {Pollock}},\ }\href {\doibase 10.1126/science.1182848} {\bibfield  {journal}
  {\bibinfo  {journal} {Science}\ }\textbf {\bibinfo {volume} {328}},\ \bibinfo
  {pages} {986} (\bibinfo {year} {2010})}\BibitemShut {NoStop}%
\bibitem [{\citenamefont {Friedrich}\ and\ \citenamefont
  {Mordike}(2006)}]{MgSpringer2006}%
  \BibitemOpen
  \bibfield  {author} {\bibinfo {author} {\bibfnamefont {H.~E.}\ \bibnamefont
  {Friedrich}}\ and\ \bibinfo {author} {\bibfnamefont {B.~L.}\ \bibnamefont
  {Mordike}},\ }\href@noop {} {\emph {\bibinfo {title} {Magnesium
  Technology}}},\ \bibinfo {edition} {1st}\ ed.\ (\bibinfo  {publisher}
  {Springer-Verlag Berlin Heidelberg},\ \bibinfo {year} {2006})\BibitemShut
  {NoStop}%
\bibitem [{\citenamefont {Joost}(2012)}]{Joost2012}%
  \BibitemOpen
  \bibfield  {author} {\bibinfo {author} {\bibfnamefont {W.}~\bibnamefont
  {Joost}},\ }\href {\doibase 10.1007/s11837-012-0424-z} {\bibfield  {journal}
  {\bibinfo  {journal} {JOM}\ }\textbf {\bibinfo {volume} {64}},\ \bibinfo
  {pages} {1032} (\bibinfo {year} {2012})}\BibitemShut {NoStop}%
\bibitem [{\citenamefont {Fromm}\ and\ \citenamefont
  {H\"orz}(1980)}]{Fromm1980}%
  \BibitemOpen
  \bibfield  {author} {\bibinfo {author} {\bibfnamefont {E.}~\bibnamefont
  {Fromm}}\ and\ \bibinfo {author} {\bibfnamefont {G.}~\bibnamefont {H\"orz}},\
  }\href {\doibase 10.1179/imtr.1980.25.1.269} {\bibfield  {journal} {\bibinfo
  {journal} {International Metals Reviews}\ }\textbf {\bibinfo {volume} {25}},\
  \bibinfo {pages} {269} (\bibinfo {year} {1980})}\BibitemShut {NoStop}%
\bibitem [{\citenamefont {Wu}\ \emph {et~al.}(2013)\citenamefont {Wu},
  \citenamefont {Liu}, \citenamefont {Wang}, \citenamefont {Gan},\ and\
  \citenamefont {Liu}}]{Wu2013}%
  \BibitemOpen
  \bibfield  {author} {\bibinfo {author} {\bibfnamefont {X.-Z.}\ \bibnamefont
  {Wu}}, \bibinfo {author} {\bibfnamefont {L.-L.}\ \bibnamefont {Liu}},
  \bibinfo {author} {\bibfnamefont {R.}~\bibnamefont {Wang}}, \bibinfo {author}
  {\bibfnamefont {L.-Y.}\ \bibnamefont {Gan}}, \ and\ \bibinfo {author}
  {\bibfnamefont {Q.}~\bibnamefont {Liu}},\ }\href {\doibase
  10.1007/s11706-013-0221-9} {\bibfield  {journal} {\bibinfo  {journal}
  {Frontiers of Materials Science}\ }\textbf {\bibinfo {volume} {7}},\ \bibinfo
  {pages} {405} (\bibinfo {year} {2013})}\BibitemShut {NoStop}%
\bibitem [{\citenamefont {Bertin}\ \emph {et~al.}(1980)\citenamefont {Bertin},
  \citenamefont {Parisot},\ and\ \citenamefont {Gacougnolle}}]{Bertin1980}%
  \BibitemOpen
  \bibfield  {author} {\bibinfo {author} {\bibfnamefont {Y.}~\bibnamefont
  {Bertin}}, \bibinfo {author} {\bibfnamefont {J.}~\bibnamefont {Parisot}}, \
  and\ \bibinfo {author} {\bibfnamefont {J.}~\bibnamefont {Gacougnolle}},\
  }\href {\doibase http://dx.doi.org/10.1016/0022-5088(80)90049-1} {\bibfield
  {journal} {\bibinfo  {journal} {Journal of the Less Common Metals}\ }\textbf
  {\bibinfo {volume} {69}},\ \bibinfo {pages} {121 } (\bibinfo {year}
  {1980})}\BibitemShut {NoStop}%
\bibitem [{\citenamefont {Wu}\ and\ \citenamefont {Trinkle}(2011)}]{Wu2011}%
  \BibitemOpen
  \bibfield  {author} {\bibinfo {author} {\bibfnamefont {H.~H.}\ \bibnamefont
  {Wu}}\ and\ \bibinfo {author} {\bibfnamefont {D.~R.}\ \bibnamefont
  {Trinkle}},\ }\href {\doibase 10.1103/PhysRevLett.107.045504} {\bibfield
  {journal} {\bibinfo  {journal} {Phys. Rev. Lett.}\ }\textbf {\bibinfo
  {volume} {107}},\ \bibinfo {pages} {045504} (\bibinfo {year}
  {2011})}\BibitemShut {NoStop}%
\bibitem [{\citenamefont {O'Hara}\ and\ \citenamefont
  {Demkov}(2014)}]{Hara2014}%
  \BibitemOpen
  \bibfield  {author} {\bibinfo {author} {\bibfnamefont {A.}~\bibnamefont
  {O'Hara}}\ and\ \bibinfo {author} {\bibfnamefont {A.~A.}\ \bibnamefont
  {Demkov}},\ }\href {\doibase http://dx.doi.org/10.1063/1.4880657} {\bibfield
  {journal} {\bibinfo  {journal} {Applied Physics Letters}\ }\textbf {\bibinfo
  {volume} {104}},\ \bibinfo {eid} {211909} (\bibinfo {year}
  {2014})}\BibitemShut {NoStop}%
\bibitem [{\citenamefont {Wu}\ \emph {et~al.}(2016)\citenamefont {Wu},
  \citenamefont {Wisesa},\ and\ \citenamefont {Trinkle}}]{Wu2016}%
  \BibitemOpen
  \bibfield  {author} {\bibinfo {author} {\bibfnamefont {H.~H.}\ \bibnamefont
  {Wu}}, \bibinfo {author} {\bibfnamefont {P.}~\bibnamefont {Wisesa}}, \ and\
  \bibinfo {author} {\bibfnamefont {D.~R.}\ \bibnamefont {Trinkle}},\ }\href
  {\doibase 10.1103/PhysRevB.94.014307} {\bibfield  {journal} {\bibinfo
  {journal} {Phys. Rev. B}\ }\textbf {\bibinfo {volume} {94}},\ \bibinfo
  {pages} {014307} (\bibinfo {year} {2016})}\BibitemShut {NoStop}%
\bibitem [{\citenamefont {Zotov}\ and\ \citenamefont
  {Tseldkin}(1976)}]{Zotov1976}%
  \BibitemOpen
  \bibfield  {author} {\bibinfo {author} {\bibfnamefont {V.}~\bibnamefont
  {Zotov}}\ and\ \bibinfo {author} {\bibfnamefont {A.}~\bibnamefont
  {Tseldkin}},\ }\href@noop {} {\bibfield  {journal} {\bibinfo  {journal}
  {Soviet Physics Journal}\ }\textbf {\bibinfo {volume} {19}},\ \bibinfo
  {pages} {1652} (\bibinfo {year} {1976})}\BibitemShut {NoStop}%
\bibitem [{\citenamefont
  {Trinkle}(2016{\natexlab{a}})}]{TrinkleElastodiffusivity2016}%
  \BibitemOpen
  \bibfield  {author} {\bibinfo {author} {\bibfnamefont {D.~R.}\ \bibnamefont
  {Trinkle}},\ }\href {\doibase 10.1080/14786435.2016.1212175} {\bibfield
  {journal} {\bibinfo  {journal} {Philos. Mag.}\ } (\bibinfo {year}
  {2016}{\natexlab{a}}),\ 10.1080/14786435.2016.1212175}\BibitemShut {NoStop}%
\bibitem [{\citenamefont {Trinkle}(2016{\natexlab{b}})}]{OnsagerCalc}%
  \BibitemOpen
  \bibfield  {author} {\bibinfo {author} {\bibfnamefont {D.~R.}\ \bibnamefont
  {Trinkle}},\ }\href {\doibase 10.5281/zenodo.57407} {\enquote {\bibinfo
  {title} {\textsc{Onsager}},}\ }\bibinfo {howpublished}
  {http://dallastrinkle.github.io/Onsager} (\bibinfo {year}
  {2016}{\natexlab{b}})\BibitemShut {NoStop}%
\bibitem [{\citenamefont {Dederichs}\ and\ \citenamefont
  {Schroeder}(1978)}]{Dederichs1978}%
  \BibitemOpen
  \bibfield  {author} {\bibinfo {author} {\bibfnamefont {P.~H.}\ \bibnamefont
  {Dederichs}}\ and\ \bibinfo {author} {\bibfnamefont {K.}~\bibnamefont
  {Schroeder}},\ }\href {\doibase 10.1103/PhysRevB.17.2524} {\bibfield
  {journal} {\bibinfo  {journal} {Phys. Rev. B}\ }\textbf {\bibinfo {volume}
  {17}},\ \bibinfo {pages} {2524} (\bibinfo {year} {1978})}\BibitemShut
  {NoStop}%
\bibitem [{\citenamefont {Savino}\ and\ \citenamefont
  {Smetniansky-De~Grande}(1987)}]{Savino1987}%
  \BibitemOpen
  \bibfield  {author} {\bibinfo {author} {\bibfnamefont {E.~J.}\ \bibnamefont
  {Savino}}\ and\ \bibinfo {author} {\bibfnamefont {N.}~\bibnamefont
  {Smetniansky-De~Grande}},\ }\href {\doibase 10.1103/PhysRevB.35.6064}
  {\bibfield  {journal} {\bibinfo  {journal} {Phys. Rev. B}\ }\textbf {\bibinfo
  {volume} {35}},\ \bibinfo {pages} {6064} (\bibinfo {year}
  {1987})}\BibitemShut {NoStop}%
\bibitem [{\citenamefont {Woo}\ and\ \citenamefont {So}(2000)}]{Woo2000}%
  \BibitemOpen
  \bibfield  {author} {\bibinfo {author} {\bibfnamefont {C.~H.}\ \bibnamefont
  {Woo}}\ and\ \bibinfo {author} {\bibfnamefont {C.~B.}\ \bibnamefont {So}},\
  }\href {\doibase 10.1080/01418610008212120} {\bibfield  {journal} {\bibinfo
  {journal} {Philosophical Magazine A}\ }\textbf {\bibinfo {volume} {80}},\
  \bibinfo {pages} {1299} (\bibinfo {year} {2000})}\BibitemShut {NoStop}%
\bibitem [{\citenamefont {Kresse}\ and\ \citenamefont
  {Furthm\"uller}(1996)}]{Kresse1996}%
  \BibitemOpen
  \bibfield  {author} {\bibinfo {author} {\bibfnamefont {G.}~\bibnamefont
  {Kresse}}\ and\ \bibinfo {author} {\bibfnamefont {J.}~\bibnamefont
  {Furthm\"uller}},\ }\href {\doibase 10.1103/PhysRevB.54.11169} {\bibfield
  {journal} {\bibinfo  {journal} {Phys. Rev. B}\ }\textbf {\bibinfo {volume}
  {54}},\ \bibinfo {pages} {11169} (\bibinfo {year} {1996})}\BibitemShut
  {NoStop}%
\bibitem [{\citenamefont {Bl\"ochl}(1994)}]{Blochl1994a}%
  \BibitemOpen
  \bibfield  {author} {\bibinfo {author} {\bibfnamefont {P.~E.}\ \bibnamefont
  {Bl\"ochl}},\ }\href {\doibase 10.1103/PhysRevB.50.17953} {\bibfield
  {journal} {\bibinfo  {journal} {Phys. Rev. B}\ }\textbf {\bibinfo {volume}
  {50}},\ \bibinfo {pages} {17953} (\bibinfo {year} {1994})}\BibitemShut
  {NoStop}%
\bibitem [{\citenamefont {Kresse}\ and\ \citenamefont
  {Joubert}(1999)}]{Kresse1999}%
  \BibitemOpen
  \bibfield  {author} {\bibinfo {author} {\bibfnamefont {G.}~\bibnamefont
  {Kresse}}\ and\ \bibinfo {author} {\bibfnamefont {D.}~\bibnamefont
  {Joubert}},\ }\href {\doibase 10.1103/PhysRevB.59.1758} {\bibfield  {journal}
  {\bibinfo  {journal} {Phys. Rev. B}\ }\textbf {\bibinfo {volume} {59}},\
  \bibinfo {pages} {1758} (\bibinfo {year} {1999})}\BibitemShut {NoStop}%
\bibitem [{\citenamefont {Perdew}\ \emph {et~al.}(1996)\citenamefont {Perdew},
  \citenamefont {Burke},\ and\ \citenamefont {Ernzerhof}}]{Perdew1996}%
  \BibitemOpen
  \bibfield  {author} {\bibinfo {author} {\bibfnamefont {J.~P.}\ \bibnamefont
  {Perdew}}, \bibinfo {author} {\bibfnamefont {K.}~\bibnamefont {Burke}}, \
  and\ \bibinfo {author} {\bibfnamefont {M.}~\bibnamefont {Ernzerhof}},\ }\href
  {\doibase 10.1103/PhysRevLett.77.3865} {\bibfield  {journal} {\bibinfo
  {journal} {Phys. Rev. Lett.}\ }\textbf {\bibinfo {volume} {77}},\ \bibinfo
  {pages} {3865} (\bibinfo {year} {1996})}\BibitemShut {NoStop}%
\bibitem [{\citenamefont {Methfessel}\ and\ \citenamefont
  {Paxton}(1989)}]{Methfessel1989}%
  \BibitemOpen
  \bibfield  {author} {\bibinfo {author} {\bibfnamefont {M.}~\bibnamefont
  {Methfessel}}\ and\ \bibinfo {author} {\bibfnamefont {A.~T.}\ \bibnamefont
  {Paxton}},\ }\href {\doibase 10.1103/PhysRevB.40.3616} {\bibfield  {journal}
  {\bibinfo  {journal} {Phys. Rev. B}\ }\textbf {\bibinfo {volume} {40}},\
  \bibinfo {pages} {3616} (\bibinfo {year} {1989})}\BibitemShut {NoStop}%
\bibitem [{\citenamefont {Friis}\ \emph {et~al.}(2003)\citenamefont {Friis},
  \citenamefont {Madsen}, \citenamefont {Larsen}, \citenamefont {Jiang},
  \citenamefont {Marthinsen},\ and\ \citenamefont {Holmestad}}]{Friis2003}%
  \BibitemOpen
  \bibfield  {author} {\bibinfo {author} {\bibfnamefont {J.}~\bibnamefont
  {Friis}}, \bibinfo {author} {\bibfnamefont {G.~K.~H.}\ \bibnamefont
  {Madsen}}, \bibinfo {author} {\bibfnamefont {F.~K.}\ \bibnamefont {Larsen}},
  \bibinfo {author} {\bibfnamefont {B.}~\bibnamefont {Jiang}}, \bibinfo
  {author} {\bibfnamefont {K.}~\bibnamefont {Marthinsen}}, \ and\ \bibinfo
  {author} {\bibfnamefont {R.}~\bibnamefont {Holmestad}},\ }\href {\doibase
  http://dx.doi.org/10.1063/1.1622656} {\bibfield  {journal} {\bibinfo
  {journal} {The Journal of Chemical Physics}\ }\textbf {\bibinfo {volume}
  {119}},\ \bibinfo {pages} {11359} (\bibinfo {year} {2003})}\BibitemShut
  {NoStop}%
\bibitem [{\citenamefont {Henkelman}\ \emph {et~al.}(2000)\citenamefont
  {Henkelman}, \citenamefont {Uberuaga},\ and\ \citenamefont
  {J{\'o}nsson}}]{Henkelman2000}%
  \BibitemOpen
  \bibfield  {author} {\bibinfo {author} {\bibfnamefont {G.}~\bibnamefont
  {Henkelman}}, \bibinfo {author} {\bibfnamefont {B.~P.}\ \bibnamefont
  {Uberuaga}}, \ and\ \bibinfo {author} {\bibfnamefont {H.}~\bibnamefont
  {J{\'o}nsson}},\ }\href {\doibase http://dx.doi.org/10.1063/1.1329672}
  {\bibfield  {journal} {\bibinfo  {journal} {The Journal of Chemical Physics}\
  }\textbf {\bibinfo {volume} {113}},\ \bibinfo {pages} {9901} (\bibinfo {year}
  {2000})}\BibitemShut {NoStop}%
\bibitem [{Note1()}]{Note1}%
  \BibitemOpen
  \bibinfo {note} {This approximation introduces at most a 40\%\ error in the
  attempt frequencies; the error is estimated by comparing with a large Mg
  supercell using bulk force constants, and introducing the interstitial-Mg
  force constants from the finite displacement calculations.}\BibitemShut
  {Stop}%
\bibitem [{\citenamefont {Vineyard}(1957)}]{Vineyard1957}%
  \BibitemOpen
  \bibfield  {author} {\bibinfo {author} {\bibfnamefont {G.~H.}\ \bibnamefont
  {Vineyard}},\ }\href {\doibase
  http://dx.doi.org/10.1016/0022-3697(57)90059-8} {\bibfield  {journal}
  {\bibinfo  {journal} {Journal of Physics and Chemistry of Solids}\ }\textbf
  {\bibinfo {volume} {3}},\ \bibinfo {pages} {121 } (\bibinfo {year}
  {1957})}\BibitemShut {NoStop}%
\bibitem [{Note2()}]{Note2}%
  \BibitemOpen
  \bibinfo {note} {Following Varvenne~\protect \textit {et al.}\cite
  {Varvenne2013}, we can estimate the finite-size error from using a $4\times
  4\times 3$ cell from the elastic dipoles (c.f., Table~\ref {tab:Sitedipole})
  and elastic constants. The largest (estimated) error in site
  energies---relative to the ground state---are 80 meV for B (dh), 60 meV for C
  (dh), 20 meV for N (h), and 3 meV for O (t).}\BibitemShut {Stop}%
\bibitem [{\citenamefont {Agarwal}\ and\ \citenamefont
  {Trinkle}(2016)}]{ComputationalData}%
  \BibitemOpen
  \bibfield  {author} {\bibinfo {author} {\bibfnamefont {R.}~\bibnamefont
  {Agarwal}}\ and\ \bibinfo {author} {\bibfnamefont {D.~R.}\ \bibnamefont
  {Trinkle}},\ }\href {http://hdl.handle.net/11256/694} {\enquote {\bibinfo
  {title} {Data citation: Light element diffusion in mg using first principles
  calculations: Anisotropy and elastodiffusion},}\ } (\bibinfo {year} {2016}),\
  \bibinfo {note} {http://hdl.handle.net/11256/694}\BibitemShut {NoStop}%
\bibitem [{\citenamefont {Clouet}\ \emph {et~al.}(2008)\citenamefont {Clouet},
  \citenamefont {Garruchet}, \citenamefont {Nguyen}, \citenamefont {Perez},\
  and\ \citenamefont {Becquart}}]{Clouet2008}%
  \BibitemOpen
  \bibfield  {author} {\bibinfo {author} {\bibfnamefont {E.}~\bibnamefont
  {Clouet}}, \bibinfo {author} {\bibfnamefont {S.}~\bibnamefont {Garruchet}},
  \bibinfo {author} {\bibfnamefont {H.}~\bibnamefont {Nguyen}}, \bibinfo
  {author} {\bibfnamefont {M.}~\bibnamefont {Perez}}, \ and\ \bibinfo {author}
  {\bibfnamefont {C.~S.}\ \bibnamefont {Becquart}},\ }\href {\doibase
  http://dx.doi.org/10.1016/j.actamat.2008.03.024} {\bibfield  {journal}
  {\bibinfo  {journal} {Acta Materialia}\ }\textbf {\bibinfo {volume} {56}},\
  \bibinfo {pages} {3450 } (\bibinfo {year} {2008})}\BibitemShut {NoStop}%
\bibitem [{\citenamefont {Mehrer}(2011)}]{Mehrer2011}%
  \BibitemOpen
  \bibfield  {author} {\bibinfo {author} {\bibfnamefont {H.}~\bibnamefont
  {Mehrer}},\ }\href {\doibase 10.4028/www.scientific.net/DDF.309-310.91}
  {\bibfield  {journal} {\bibinfo  {journal} {Defect and Diffusion Forum}\
  }\textbf {\bibinfo {volume} {309-310}},\ \bibinfo {pages} {91} (\bibinfo
  {year} {2011})}\BibitemShut {NoStop}%
\bibitem [{\citenamefont {Greeff}\ and\ \citenamefont
  {Moriarty}(1999)}]{Greeff1999}%
  \BibitemOpen
  \bibfield  {author} {\bibinfo {author} {\bibfnamefont {C.~W.}\ \bibnamefont
  {Greeff}}\ and\ \bibinfo {author} {\bibfnamefont {J.~A.}\ \bibnamefont
  {Moriarty}},\ }\href {\doibase 10.1103/PhysRevB.59.3427} {\bibfield
  {journal} {\bibinfo  {journal} {Phys. Rev. B}\ }\textbf {\bibinfo {volume}
  {59}},\ \bibinfo {pages} {3427} (\bibinfo {year} {1999})}\BibitemShut
  {NoStop}%
\bibitem [{\citenamefont {Wuttig}\ and\ \citenamefont
  {Keiser}(1971)}]{Wuttig1971}%
  \BibitemOpen
  \bibfield  {author} {\bibinfo {author} {\bibfnamefont {M.}~\bibnamefont
  {Wuttig}}\ and\ \bibinfo {author} {\bibfnamefont {J.}~\bibnamefont
  {Keiser}},\ }\href {\doibase 10.1103/PhysRevB.3.815} {\bibfield  {journal}
  {\bibinfo  {journal} {Phys. Rev. B}\ }\textbf {\bibinfo {volume} {3}},\
  \bibinfo {pages} {815} (\bibinfo {year} {1971})}\BibitemShut {NoStop}%
\bibitem [{\citenamefont {Bosman}\ \emph {et~al.}(1960)\citenamefont {Bosman},
  \citenamefont {Brommer}, \citenamefont {Eijkelenboom}, \citenamefont
  {Schinkel},\ and\ \citenamefont {Rathenau}}]{Bosman1960}%
  \BibitemOpen
  \bibfield  {author} {\bibinfo {author} {\bibfnamefont {A.}~\bibnamefont
  {Bosman}}, \bibinfo {author} {\bibfnamefont {P.}~\bibnamefont {Brommer}},
  \bibinfo {author} {\bibfnamefont {L.}~\bibnamefont {Eijkelenboom}}, \bibinfo
  {author} {\bibfnamefont {C.}~\bibnamefont {Schinkel}}, \ and\ \bibinfo
  {author} {\bibfnamefont {G.}~\bibnamefont {Rathenau}},\ }\href {\doibase
  http://dx.doi.org/10.1016/0031-8914(60)90105-1} {\bibfield  {journal}
  {\bibinfo  {journal} {Physica}\ }\textbf {\bibinfo {volume} {26}},\ \bibinfo
  {pages} {533 } (\bibinfo {year} {1960})}\BibitemShut {NoStop}%
\bibitem [{\citenamefont {Touloukian}\ \emph {et~al.}(1975)\citenamefont
  {Touloukian}, \citenamefont {Kirby}, \citenamefont {Taylor},\ and\
  \citenamefont {Lee}}]{TEdata}%
  \BibitemOpen
  \bibfield  {author} {\bibinfo {author} {\bibfnamefont {Y.}~\bibnamefont
  {Touloukian}}, \bibinfo {author} {\bibfnamefont {R.}~\bibnamefont {Kirby}},
  \bibinfo {author} {\bibfnamefont {R.}~\bibnamefont {Taylor}}, \ and\ \bibinfo
  {author} {\bibfnamefont {T.}~\bibnamefont {Lee}},\ }\href@noop {} {\emph
  {\bibinfo {title} {Thermophysical Properties of Matter: Thermal
  Expansion-Metallic Elements and Alloys}}},\ \bibinfo {edition} {1st}\ ed.,\
  Vol.~\bibinfo {volume} {12}\ (\bibinfo  {publisher} {Plenum Press, New
  York},\ \bibinfo {year} {1975})\BibitemShut {NoStop}%
\bibitem [{Jmo()}]{Jmol}%
  \BibitemOpen
  \href {http://www.jmol.org/} {}\bibinfo {note} {Jmol: an open-source Java
  viewer for chemical structures in 3D}\BibitemShut {NoStop}%
\bibitem [{\citenamefont {Varvenne}\ \emph {et~al.}(2013)\citenamefont
  {Varvenne}, \citenamefont {Bruneval}, \citenamefont {Marinica},\ and\
  \citenamefont {Clouet}}]{Varvenne2013}%
  \BibitemOpen
  \bibfield  {author} {\bibinfo {author} {\bibfnamefont {C.}~\bibnamefont
  {Varvenne}}, \bibinfo {author} {\bibfnamefont {F.}~\bibnamefont {Bruneval}},
  \bibinfo {author} {\bibfnamefont {M.-C.}\ \bibnamefont {Marinica}}, \ and\
  \bibinfo {author} {\bibfnamefont {E.}~\bibnamefont {Clouet}},\ }\href
  {\doibase 10.1103/PhysRevB.88.134102} {\bibfield  {journal} {\bibinfo
  {journal} {Phys. Rev. B}\ }\textbf {\bibinfo {volume} {88}},\ \bibinfo
  {pages} {134102} (\bibinfo {year} {2013})}\BibitemShut {NoStop}%
\end{thebibliography}
\end{document}